\documentclass[epj]{svjour}
\usepackage{graphics}
\usepackage{amsmath}
\usepackage{amsfonts}
\usepackage{amssymb}
\usepackage[pdftex]{graphicx}
\usepackage{verbatim}
\usepackage{color}
\usepackage[usenames,dvipsnames]{xcolor} 

\begin{document}
\title{Towards real-world complexity: an introduction to multiplex networks}
\author{Kyu-Min Lee\inst{1}, Byungjoon Min\inst{1}\thanks{{Present address:} Department of Physics and Levich Institute, City College of New York, New York, NY 10031, USA}, and K.-I. Goh\inst{1}\thanks{
e-mail: {\tt kgoh@korea.ac.kr}}
}                     
%
%
\institute{$^1$Department of Physics and Institute of Basic Science, Korea University, Seoul 136-713, Korea} 
\date{Received:  / Revised version:  }

\abstract{
Many real-world complex systems are best modeled by multiplex networks of interacting network layers. The multiplex network study is one of the newest and hottest themes in the statistical physics of complex networks. Pioneering studies have proven that the multiplexity has broad impact on the system's structure and function. In this Colloquium paper, we present an organized review of the growing body of current literature on multiplex networks by categorizing existing studies broadly according to the type of layer coupling in the problem. Major recent advances in the field are surveyed and some outstanding open challenges and future perspectives will be proposed.
} 

\maketitle

\section{Introduction}
\label{intro}
Many real-world complex systems ranging from living organisms and human society to transportation system and critical infrastructure operate through multiple layers of distinct interactions among constituents as well as the interplay between these interaction layers to fulfill their emergent function~\cite{NoNBook,Kivela2014,Boccaletti2014}. 
Multiplex network \cite{Mucha2010,KMLee2012,Brummitt2012,Gomez2013,Bianconi2013,Kim2013,Min2014b,Baxter2014} is a class of networks introduced to better model such systems, in which the same set of nodes are connected via more than one type of links\footnote{In the literature, the term ``multiplex'' tends to be used in a more loose sense, referring also to more general multilayered and interconnected systems. Throughout this paper, however, we will try to adhere to the strict definition of the term.} (Fig.~1). Each type of links in multiplex networks constitutes the network layer. 
Examples for multiplex networks abound: Individuals in a society are networking through numerous social relationships such as friendship, kinship, co-workership and via a multitude of communication channels such as online and offline contacts \cite{Wasserman1994}. 
Critical infrastructure provides essential support for the functioning of modern society through concerted operations of multiple interlinked and interrelated networks such as energy production and supply, telecommunication, and transportation networks \cite{Nagurney2009}.
The study of multiplex networks has emerged as one of the major contemporary topics of network theory \cite{KMLee2012,Brummitt2012,Gomez2013,Bianconi2013,Kim2013,Min2014b,Baxter2014}, along with the parallel development of closely related topics such as interacting networks \cite{Leicht2009}, interdependent networks \cite{Buldyrev2010}, and interconnected networks \cite{Radicchi2013}.

\begin{figure}
\centering
\includegraphics[width= .75\columnwidth]{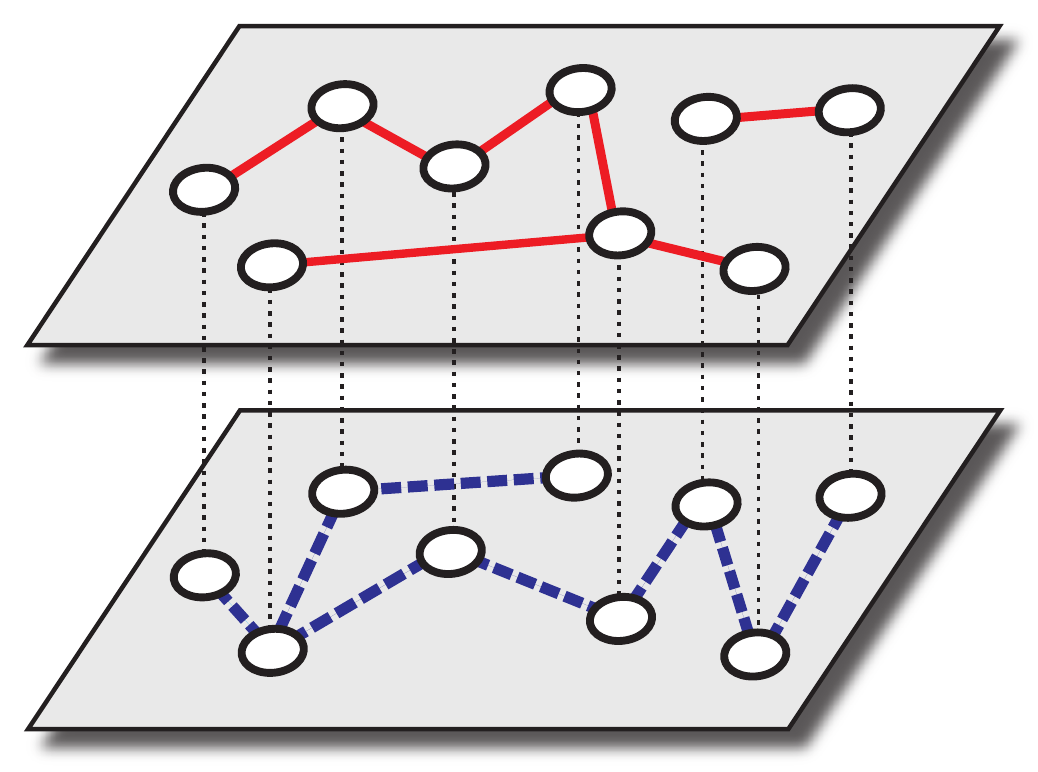}
\caption{An illustrative example of the multiplex network of nine nodes with two layers, the red (solid) and the blue (dashed) layer. A node and its replica are shown linked by the dotted line, to denote the identity relation. In addition to the multiplex network topology like this, the type of layer coupling needs also to be implemented correctly, for a reliable treatment of given multiplex problem. 
}
\label{fig:functional-coupling}
\end{figure}

Introduction of manifold layers in multiplex networks necessitates new conceptual as well as computational developments beyond the standard single-network theoretic framework that has been well-established during the last fifteen or so years \cite{NewmanBook}. The multiplex network is not just a simple stack of many network layers. 
What makes a truly multiplex system is the functional coupling between layers, which gives rise to in general non-additive and nonlinear effect for the emergent function of the system.
Hence a proper consideration of the contexts of multiplexity that the multiple layers make is essential for correct description and successful understanding of multiplex problems.

There are two important facets of multiplexity, which we might refer to as (i) the pattern of multiplexity and (ii) the type of multiplexity.
The former concerns how the layers are coupled {\em structurally}.
The same set of network layers can be coupled in many ways to form different multiplex structures (and {\it vice versa}). 
In real-world multiplex systems, the coupling structure between layers are far from random but correlated, a property termed the correlated multiplexity \cite{KMLee2012}, which has been assessed to be significant empirically \cite{Szell2010} in terms of the interlayer degree correlation and the link overlap. 
The type of multiplexity concerns how the layers are coupled {\em functionally}, that is, how the function of one layer affects that of another (and {\it vice versa}). For instance, the multiple layers in a multiplex system may be coupled either in complementary/connective manner as in the transportation system~\cite{KMLee2012,Gomez2013,Leicht2009} or in cooperative/dependent way as in the critical infrastructure~\cite{Min2014b,Buldyrev2010}. Depending on the type of multiplexity, same multiplex structures can behave quite differently. A legitimate understanding of multiplex network problems can thus be achieved only when both structural and functional aspects of multiplexity are correctly taken into account.

In this Colloquium paper, we attempt an organic \\overview of the fast-growing body of current literature on statistical physics of multiplex networks. We will pay particular attention to put the diverse approaches in related problems into a coherent setting characterized and categorized broadly by the type of multiplexity. 
In so doing, we hope this paper not to reiterate but to complement several existing review articles on closely-related topics. Readers are referred to references~\cite{Vespignani2010,Gao2012a} for more focused summary on interdependent networks, to references~\cite{Kivela2014,Boccaletti2014} for broader expositions on multilayer networks, and to references~\cite{Holme2012,HolmeBook} for comprehensive reviews on temporal networks.
A compendium volume \cite{NoNBook} collecting contributions from pioneering groups is also available, providing an overarching overview of the early development of the field at large. 

The rest of this Colloquium paper is organized as follows.  In Section\ 2, we discuss how to characterize and model the multiplex systems. There will also be introduced the basic definitions and terminology. In Section\ 3,  we review multiplex problems with cooperatively-coupled layers, such as mutual percolation and multiplex cascades, for which major new concepts and analytic methods specific to multiplex systems are introduced. In Section\ 4, problems with complementary layers, such as  transport and  epidemic spreading on interconnected layers, are surveyed. In Section\ 5, other types of layer coupling such as competitive coupling and directional coupling are briefly discussed. The paper will close with  conclusion and outlook in Section\ 6.

\section{Characterization and modeling of the multiplex structure}

A first step towards understanding of new network concept is to formulate appropriate means to characterize the structural properties. To motivate, we first look for empirical data for multiplex networks, ranging from social and man-made systems to biological systems. Basic representations of multiplex network ensemble and various structural measures are introduced. Finally, random graph models of multiplex networks will be introduced as theoretical tools for systematic and analytic understanding of the multiplex system. 

\subsection{Empirical studies and multiplex network data}
\label{subsec:data}

Empirical studies on real-world network data have been a major fuel of network science, motivating newer and more sophisticated network concepts and developments of necessary analytic tools. 

Society is a prime example of the multiplex network system. Since early social network studies, data on the social multiplex (sometimes also termed as multi-stranded) relationships had been collected through surveys and interviews, and its social implications had been discussed~\cite{Kapferer1969,Verbrugge1979,Minor1983,Padgett1993}\footnote{In the social network literature the term multiplexity has often been used to refer the degree of overlap in relations (links) across different layers.}. With the recent development of digital technology, human individual activities are being recorded in unprecedentedly high spatiotemporal resolution and in large scale. Studies exploiting this opportunity came from the analyses of the massive multiplayer online games~\cite{Szell2010,Szell2010b,sSon2012}, first quantifying the correlated multiplexity in large scale. 
Another kind of large-scale yet more accessible multiplex social network data is the communication-collaboration network between scientists~\cite{KMLeeBook,Menichetti2014,Domenico2013a}. 

Many man-made systems in modern society are also multiplex network systems. An example is the transportation system. In the scale of urban and nation-wide transit system, several datasets on different transportation modes~\cite{Domenico2014d,Gu2011} have been constructed. At a larger scale, a multiplex European airway network with thirty-seven layers corresponding to routes operated by different commercial airline companies~\cite{Cardillo2013a} and the worldwide airport--seaport coupled network~\cite{Parshani2010} have been analyzed. 
Another example is the so-called critical infrastructure systems ranging from the power grid and water supply networks to wired and wireless information communication infrastructures~\cite{Little2002,Rosato2008}, which motivated the concept of interdependency between layers in a multiplex system, leading to interesting new physics (see Sect.~3 for more details).
The global economic system is also a multiplex system consisting of various financial and political layers. Best characterized example is the international trade network among countries as a multiplex network with layers representing different types of goods~\cite{Barigozzi2010,Foschi2014}. 

Biology is governed by multiplex networks across many levels of hierarchy. 
At the cellular level, the basic cellular function is fulfilled by the coordinated actions of many layers of biomolecular interactions. 
Reconstruction and interrogation of major biomolecular network layers in model organisms \cite{Pneumoniae1,Pneumoniae2,Pneumoniae3} and human \cite{Vidal,Palsson} have made steady progress in the last decade towards the whole-cell modeling~\cite{Karr2012}. At the physiological level, an example is provided by the neuronal network of {\it Caenorhabditis elegans} consisting of two layers of connections, the synaptic connections and the gap junctions~\cite{White1986}, 
and the layer-level structural properties have been analyzed in reference~\cite{Nicosia2014a}.
At the ecological level, species interact with one another through mutualistic, host-parasite or predator-prey relationship, which should be treated in terms of the multiplex network~\cite{Pocock2012}. 

\subsection{Mathematical formulations and measures}
\label{formula}

On first thought, it might seem straightforward to generalize various network concepts and measures well-defined for single-layer networks~\cite{NewmanBook} into multiplex networks by introducing additional index for the network layer. This turned out not always to be the case~\cite{Battiston2014}. Often such simple-minded generalization could miss the essential feature of the multiplex system, that is the context of multiplexity. In this section, we shall try to summarize the current status of theoretical development in this direction, clarifying what has been done and what has to be done. For the sake of simplicity of presentation, we shall consider only the multiplex networks consisting of layers of simple undirected unweighted graphs.

Throughout this paper, we will use the term simple multiplex to refer the multiplex networks in which a node participating to multiple layers (to be called a multiplex node) should participate to all the layers. This restriction is relaxed in more general multilayer interconnected networks. 
When every node in the network is multiplexed, the system is said to be fully multiplexed. Otherwise, it is called partially multiplex. 

\subsubsection{Matrix representation}

One can fully represent the given instance of network (or graph) with the adjacency matrix $\mathbf A$ with elements $A_{ij}=1$ if the nodes $i$ and $j$ are connected by a link and $A_{ij}=0$ otherwise. A straightforward generalization of the adjacency matrix for multiplex network is the so-called supra-adjacency matrix \cite{Domenico2013b} or adjacency tensor \cite{Mucha2010}. For a multiplex network with $N$ nodes and $\ell$ layers, the supra-adjacency matrix ${\cal A}$ is an $\ell \times\ell$ matrix of $N\times N$ blocks of layer-to-layer adjacency matrices. The $(ij)$-element of $(ab)$-block, $({\cal A}_{ab})_{ij}$, for a simple multiplex is thus given by  
\begin{align}
({\cal A}_{aa})_{ij} = \left\{ \begin{array}{ll}
1, & \textrm{if nodes $i$ and $j$ are linked within layer $a$,}\\
0, & \textrm{otherwise},
\end{array} \right.
\end{align}
and 
\begin{align}
({\cal A}_{ab})_{ii}= \left\{ \begin{array}{ll}
1, & \textrm{if node $i$ is present in both layers $a$ and $b$,}\\
0, & \textrm{otherwise},
\end{array} \right.
\end{align}
with $i,j = 1,\cdots,N$, and $a,b=1,\cdots,\ell$.
Throughout this paper, the alphabetical indices $(a,b,\dots)$ will be used for the dummy index of layers, with the indices $i$ and $j$ reserved for the node index. 
In the strict definition of simple multiplex networks, the off-diagonal blocks are always diagonal matrices.
Therefore the supra-adjacency matrix representation is somewhat redundant to encode the link structure of multiplex networks, but we shall reserve this representation for its applicability to more general multilayer interconnected networks in which the interlayer links can connect different nodes across layers, resulting in nonzero off-diagonal elements in the off-diagonal blocks~\cite{Domenico2013b}. Initial idea had  been applied to study the community structure~\cite{Mucha2010}. Supra-Laplacian matrix of the multiplex network can also be defined by following the similar generalization procedure~\cite{Gomez2013}, which has been applied in formalizing  diffusion-type linear dynamic processes on multiplex networks~\cite{Gomez2013,Sole-Ribalta2013,Cozzo2013} (see Sect.~4.2 for more details) and in generalizing the eigenvector centrality~\cite{Sola2013a}.  Generalization of the celebrated PageRank algorithm to the multiplex network has also been proposed~\cite{Halu2013}. 
Despite its mathematical elegance and theoretical appeal, the supra-adjacency matrix based representation {\it per se} is not sufficient in that it does not fully encode the functional context of multiplexity, which should be specified separately. 

\begin{figure*}
\label{fig:correlated}
\centering
\includegraphics[width= .8\linewidth]{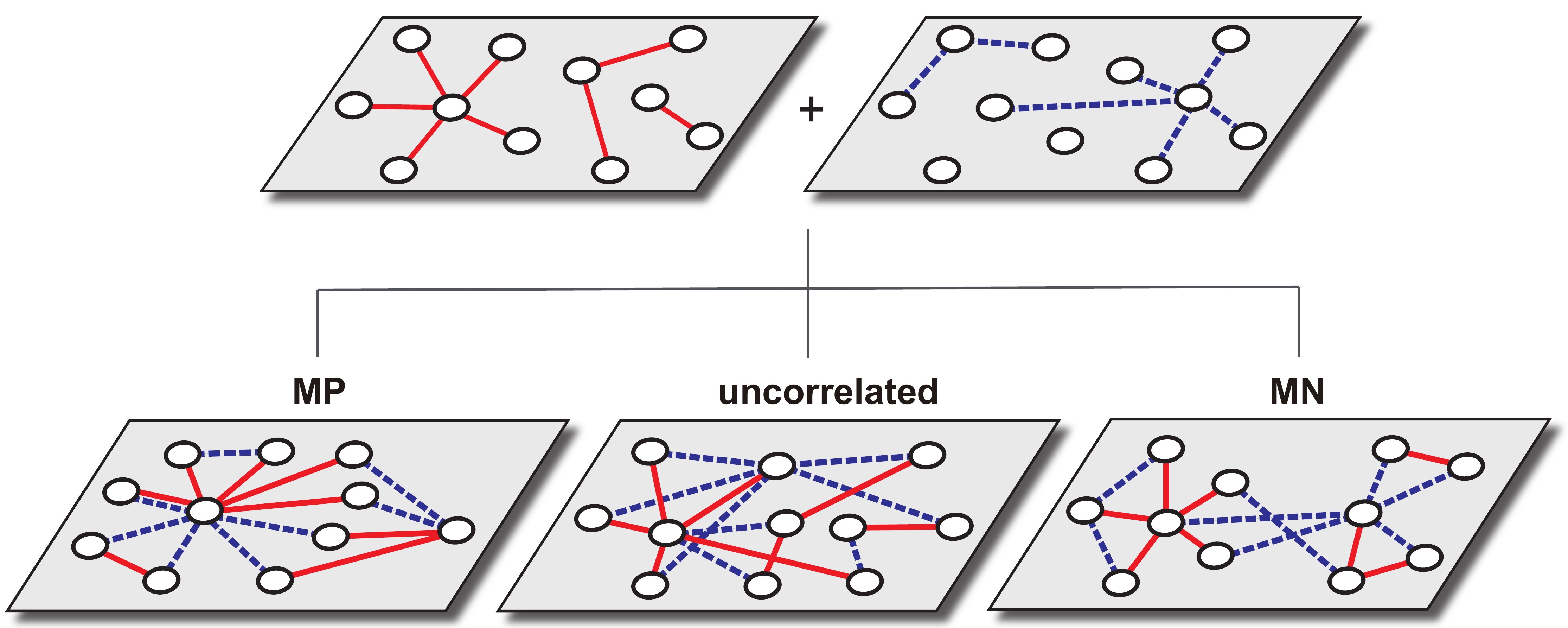}
\caption{Schematic illustrations of three patterns of interlayer degree-correlated multiplex networks proposed in \cite{KMLee2012}, the maximally positive (MP), uncorrelated, and maximally negative (MN) cases of two layers (the red solid and the blue dotted layers). 
}
\end{figure*}

\subsubsection{Multiplex degree}

The multiplex degree of node $i$ (or multidegree as is sometimes called) is constructed as a list of its degrees in each layer,
\begin{align}
{\mathbf k}_{i} = \Big\{{k_{i}^{(1)}, \dots, k_{i}^{(\ell)}}\Big\},
\end{align}
with $k_{i}^{(a)} = \sum_{j} ({\cal A}_{aa})_{ij}$, where the layer index is written superscripted to avoid confusion with the node index.
(note that in the general multilayer case, interlayer degrees $\{k_i^{(ab)}\}$ need also be specified).

The concept of degree centrality can be generalized in many ways based on the multiplex degree. Proper definition of degree centrality would depend on the system's functional characteristics and the context of multiplexity. For example, in a system with complementary layers the total degree of a node could be a good candidate for a degree centrality measure. In the presence of overlapping links, however, either the total sum of the degree in each layer or the total number of non-redundant links over all the layers could be more meaningful, depending on the physical process for which the centrality aims to address. Moreover, in systems with cooperative or competitive layers it would not even be clear whether the total degree could be a meaningful quantity for centrality at all (considering, for example, the case of mutual connectivity, which is discussed in detail in Sect.~3). In general, the concept of centrality of a node should be generalized with care in multiplex networks, since a context-blind generalization can lead to irrelevant or even misleading implications. 

The joint distribution of multiplex degree $P({\mathbf k})$ is a natural generalization of the degree distribution $P(k)$ for single-layer network, encoding fundamental structural information of a multiplex network ensemble \cite{Leicht2009}.

\subsubsection{Correlations between layers}

In multiplex networks, the correlation property of the layer coupling is an important part of the context of multiplexity, introducing extra dimension of network correlations in addition to the correlation properties of individual layers. The lowest-order layer-level correlation would be the extent of node multiplexity, {\it viz.}, what fraction of nodes in the network appear across different layers (multiplexed). 

To the next order, the interlayer degree correlation specify how correlated the degrees of a node are across different layers. It is easy to conceive the prevalence of this type of correlation in multiplex systems: a person with many friends would likely have many colleagues also in the workplace, being a friendly person; an airway-hub city tends to be very well-connected in ground as well, through railways and highways. 
In the presence of interlayer degree correlation the joint degree distribution $P({\mathbf k})$ does not factorize into the product of individual layer's degree distribution.
In parallel with the assortativity measure for single-layer network \cite{Newman2002}, simplified measures based on the correlation coefficients were introduced to quantify the interlayer degree correlation: Pearson correlation coefficient between two layers $a$ and $b$,
\begin{align}
\rho_{ab} = \frac{\langle k_{a}k_{b}\rangle-\langle k_{a}\rangle\langle k_{b}\rangle}{\sigma_{k_{a}}\sigma_{k_{b}}},
\end{align}
where $k_{a,b}$ denote the degrees of the same nodes in different layers, $\langle x\rangle$ the average over all multiplex nodes, and $\sigma_x$ the standard deviation thereof, has been used for this purpose in empirical \cite{Szell2010} as well as theoretical studies \cite{Kim2013}. 
Other correlation coefficients like Spearman rank correlation or Kendall's $\tau$ rank correlation have also been considered~\cite{Nicosia2014a}.  The average degree in layer $a$, $\langle k_a\rangle$, of nodes having the same degree $k_b$ in the other layer $b$, $\langle k_{a}\rangle(k_b)$, can also encode the interlayer degree correlation property: Increasing (decreasing) behavior of $\langle k_{a}\rangle(k_{b})$ signifies positive (negative) interlayer degree correlation \cite{Nicosia2013a}. More kinds of interlayer correlations were introduced and empirically measured in reference~\cite{Nicosia2014a}.

To investigate systematically the effect of interlayer degree correlation, it has been proposed to compare three specific patterns of correlated coupling, the maximally-positive (MP) coupling, random (uncorrelated) coupling, and the maximally-negative (MN) coupling \cite{KMLee2012} (Fig.~2). 
Given the two layers, the MP-coupled multiplex is constructed by matching the nodes of the same degree-rank from each layer. That is, the largest (lowest) degree node in one layer is coupled to the largest (lowest) degree node in the other layer. The MN coupling is constructed in the opposite way of MP coupling.  
Effects of interlayer degree correlation have been studied in this way for percolation~\cite{KMLee2012}, mutual percolation and robustness \cite{Min2014a}, and cascading failures \cite{Tan2013} problems.

Another important layer-level multiplex correlation is the presence and extent of link overlap across different layers. In social network literature, the term multiplexity often refers to this feature, asking questions like how the existence of link in one layer facilitates or constrains the formation of the link between the same node pair in another layer and what the distinctive functional role of the overlapping link is to various social dynamical processes. From theoretical point of view, the link overlap would be statistically unlikely to exist if the layers were coupled completely randomly (assuming sparseness of the layers).  Therefore the presence of link overlap signifies underlying non-randomness of layer coupling in the system. Empirical studies have shown that the link overlap between two social network layers is indeed very significant, with Jaccard index as high as $0.16$ in the example of Pardus network~\cite{Szell2010}. Functional roles of overlapping links have been studied theoretically for mutual percolation \cite{Hu2013,Cellai2013} and viability~\cite{Min2014c}, revealing the intricate role that overlapping links play in cooperatively-coupled multiplex networks. 

To a higher-order correlation, the closed three-body relation, or the triad, can measure the degree of transitivity of interactions, and has served as the basis for the notion of network clustering \cite{NewmanBook}. As triangular relations in multiplex networks can be closed not just within a layer but also across layers, generalized notion is required. For example, mutual friends of the same person can form business relationship more easily than other random pairs, through the brokering action by the mutual friend. 
The so-called cross-layered clustering coefficient has been defined \cite{Donges2011,Brodka2011,Brodka2012,Cardillo2013b} to address this new type of transitivity in multiplex networks. A family of clustering coefficients parametrized with the layer coupling strength has also been proposed~\cite{Cozzo2013b}. The functional meaning and impact of cross-layer transitivity for cooperatively or competitively coupled layers (such as mutual percolation introduced in Sect.~3) remain to be fully examined.

\subsubsection{Shortest path and other distance-based measures}

Similar conceptual complication also applies to the notion of shortest paths in multiplex networks and related quantities like pairwise distance and centrality measures like closeness, betweenness, and efficiency \cite{NewmanBook}. 
For systems with complementary/connective coupling (imagine for example the interconnected transportation network) it is relatively easy to generalize the notion in terms of the optimal paths along which one can reach from one node to another in minimum time, by incorporating the difference in link capacity (say different average speed of each transportation layer) and the layer-switching cost (say average transit time). Quantities defined along this line have been used in transport problems on multiplex networks \cite{Donges2011,Cardillo2013b,Morris2012}.
Still, the meaning and relevance of the shortest path and distance-based measures in systems with cooperatively or competitively coupled layers are much less clear (as in the case of mutual connectivity in Sect.~3). 
Proper distance measures on multiplex networks should thus be context-dependent and incorporate the functional aspect of the layer coupling and the dynamic process under consideration, calling for more attention.

\subsection{Multiplex network models}

\subsubsection{Multiplex random graph models}

Random graph models play two important roles in network theory. First they provide null models for network structure with which empirical structure of real-world networks can be compared and key structural characteristics can be extracted. Second they serve as theoretical testbeds on which systematic analytic and computational analyses can be performed to establish basic understanding of the structural and dynamical characteristics of the networks. 

Random graph ensembles are constructed by specifying structural constraints, such as the number of nodes and links, degrees of nodes, degree-degree correlations between connected nodes, {\it etc.}, as one tries to model networks with higher order correlations. 
Most elementary and widely-used graph ensemble is the one with the degree sequence or degree distribution constrained. Such an ensemble is straightforward to generalize to multiplex networks by using the multiplex degree ${\mathbf k}=\{k_a\}$ or its distribution $P({\mathbf k})$ \cite{KMLee2012,Bianconi2013,Leicht2009,Buldyrev2010,KMLeeBook}. 
(In more general multilayer case, the interlayer degrees $\{k_{ab}\}$ need also be used \cite{Bianconi2013,Leicht2009}). 
Using such random graph models, percolation problems of uncorrelated interconnected \cite{Leicht2009} and interdependent layers \cite{Buldyrev2010} were first studied. Percolation of multiplex networks with interlayer degree correlation was investigated in references~\cite{KMLee2012,KMLeeBook,Parshani2010,Min2014a,Buldyrev2011}. 
It is worthwhile to note also the related predecessors like random graphs with colored edges  \cite{Soderberg2003a,Soderberg2003b} and  the multi-component static scale-free network model~\cite{DHKim2004}.

Networks with finite fraction of overlapping links cannot be constructed in the above way. To handle this, one need to constrain the degrees for the overlapping links separately. In the case of multiplex networks with two layers, for example, this can be done by decomposing the multiplex degree as ${\mathbf k}=\{k_A,k_B,k_{\mathrm{overlap}}\}$, with $k_{\mathrm{overlap}}$ representing the degree for overlapping links and the other two for non-overlapping links. 
The random multiplex ensemble with link overlap was introduced in \cite{Bianconi2013} (therein the overlapping links were called the multilinks). Using this method the mutual percolation property of multiplexes with link overlap were investigated \cite{Hu2013,Cellai2013,Min2014c}.  
Spatially embedded multiplex ensemble was also proposed as a possible way to produce the link overlap in multiplex networks~\cite{Halu2014}.

\subsubsection{Growing multiplex network models} 
Growing network models represent another important class of network models, in which
the generative rules specifying how the new link is formed as the new nodes and links are introduced in (pseudo) time to the network define the particular random graph process. 
This class of models have been useful in providing insights on the relationship between the microscopic linkage rules and the macroscopic network properties~\cite{NewmanBook}, as in the famed example of the preferential attachment for scale-free networks~\cite{Barabasi1999}.
On this basis, there have been attempts for the multiplex network evolution model~\cite{Kim2013,Nicosia2013a,Podobnik2012}. 

A notable new idea for multiplex evolution model is that the layers coexisting in multiplex system may also coevolve. The multiplex layers do not merely evolve together in time, but the evolution of layers can also become entangled, in the sense that evolution of one layer is dependent not only on the state of the current layer but also on those of other layers. 
This idea has been implemented using the preferential attachment scheme~\cite{Kim2013,Nicosia2013a}. 
We illustrate the main idea and results following~\cite{Kim2013}. 

\begin{figure}
\centering
\includegraphics[width=.9\columnwidth]{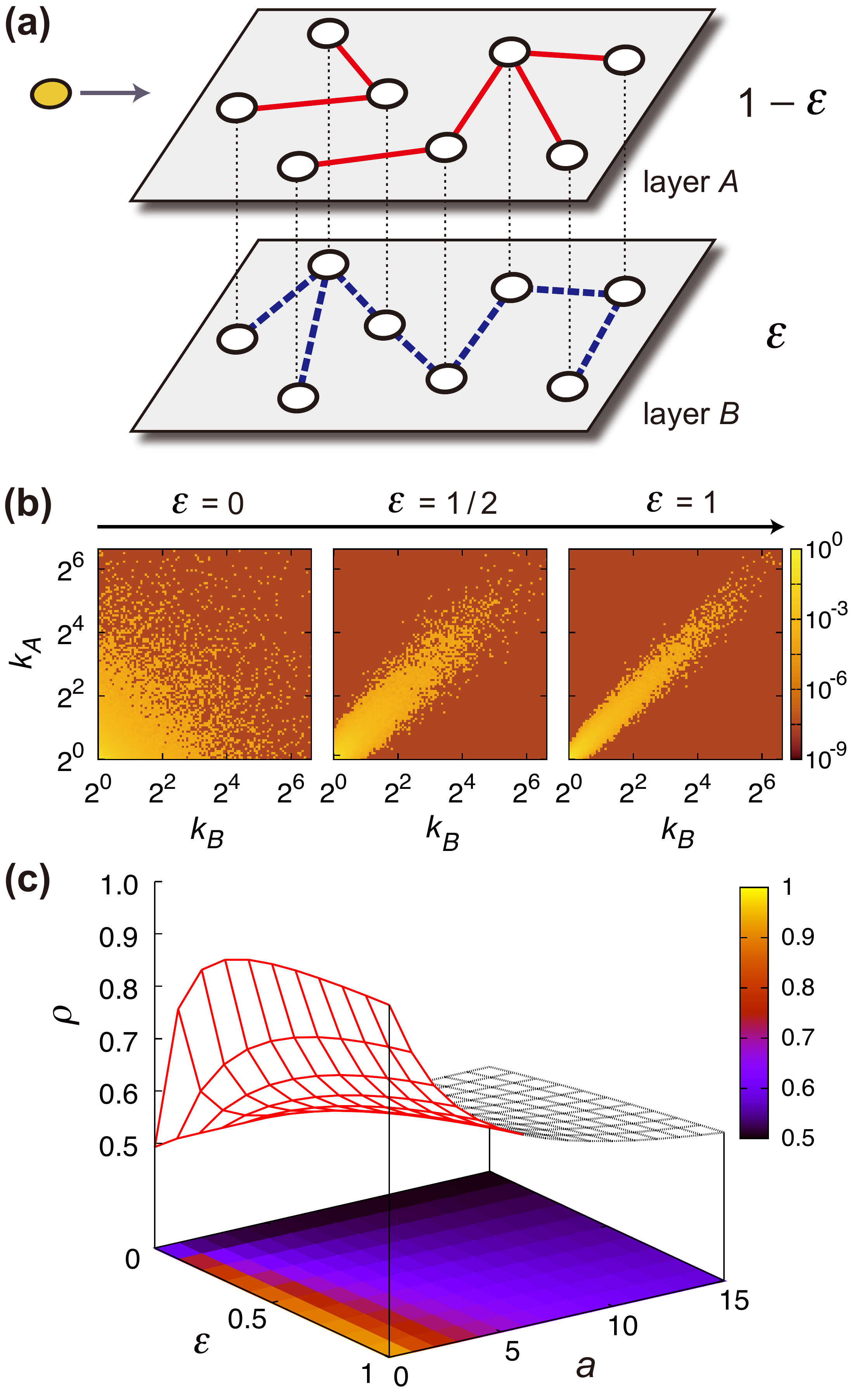}
\caption{(a) Schematic illustration for the coevolution model. Each step, a new node (orange) enters the system and establishes a link in each layer. To choose the node to connect in the layer $A$, the new node refers to the nodes' degrees not only in that layer $A$ but also in the other layer $B$ (and similarly in the layer $B$). Relative dependency to the other layer is controlled by the coevolution parameter $\epsilon$. (b) Joint degree distribution $P(k_A,k_B)$ for different coevolution parameter $\epsilon$, demonstrating the increasing interlayer degree correlation with coevolution. 
(c) Interlayer degree correlation coefficient $\rho$ as a function of coevolution parameter $\epsilon$ and shift factor $a$. Adapted from reference~\cite{Kim2013}.}
\label{fig:generating}
\end{figure}

In the coevolving multiplex network growth model, a node is newly added to the system in each step and connects to existing nodes following the growth kernel $\mathrm{\Pi}$.  
The coevolution property is implemented in the way that the growth kernel $\mathrm\Pi_{a}$ of a node's degree in layer $a$ depends not only  on its degree in that layer but also its degrees in other layers, 
\begin{align}
\mathrm{\Pi}_a = f\Big(k^{(1)},k^{(2)},\dots,k^{(a)},\dots,k^{(\ell)}\Big),
\end{align}
where the superscripted layer indices are used.
A simple example for two-layer network  (Fig.~\ref{fig:generating}a) is provided by the growth kernels in the form of coupled linear preferential attachment, given by~\cite{Kim2013}:
\begin{align}
\mathrm{\Pi}_A \propto [(1-\epsilon)(k_A+a)+\epsilon(k_B+a)],\nonumber\\
\mathrm{\Pi}_B \propto [\epsilon(k_A+a)+(1-\epsilon)(k_B+a)],
\end{align}
where the constant $\epsilon$ is the so-called coevolution parameter and $a$ is the constant determining the layer's native degree exponent. The degree distribution of each layer is shown to be independent of the value of $\epsilon$ and becomes $P(k)\sim k^{-(3+a)}$ for both layers. However, the degree correlation between layers is strongly affected by the coevolution parameter $\epsilon$ (Fig.~\ref{fig:generating}b). Specifically the interlayer degree correlation coefficient $\rho$ defined by equation~(4) is obtained asymptotically as 
\begin{align}
\rho = \frac{6\epsilon + a}{6\epsilon + 2a}
\end{align}
for $a\ge0$ \cite{Kim2013}.
The interlayer degree correlation increases with the strength of coupling between the layers' growth, quantified by the coevolution parameter $\epsilon$ in this model (Fig.~3c). 

The model proposed in~\cite{Nicosia2013a} is similar, and was later extended also to the case of nonlinear growth kernels~\cite{Nicosia2013b}, showing how the structural instability might emerge  in growing multiplex networks.

\begin{figure*}
\label{fig:mutual}
\centering
\includegraphics[width= .83\textwidth]{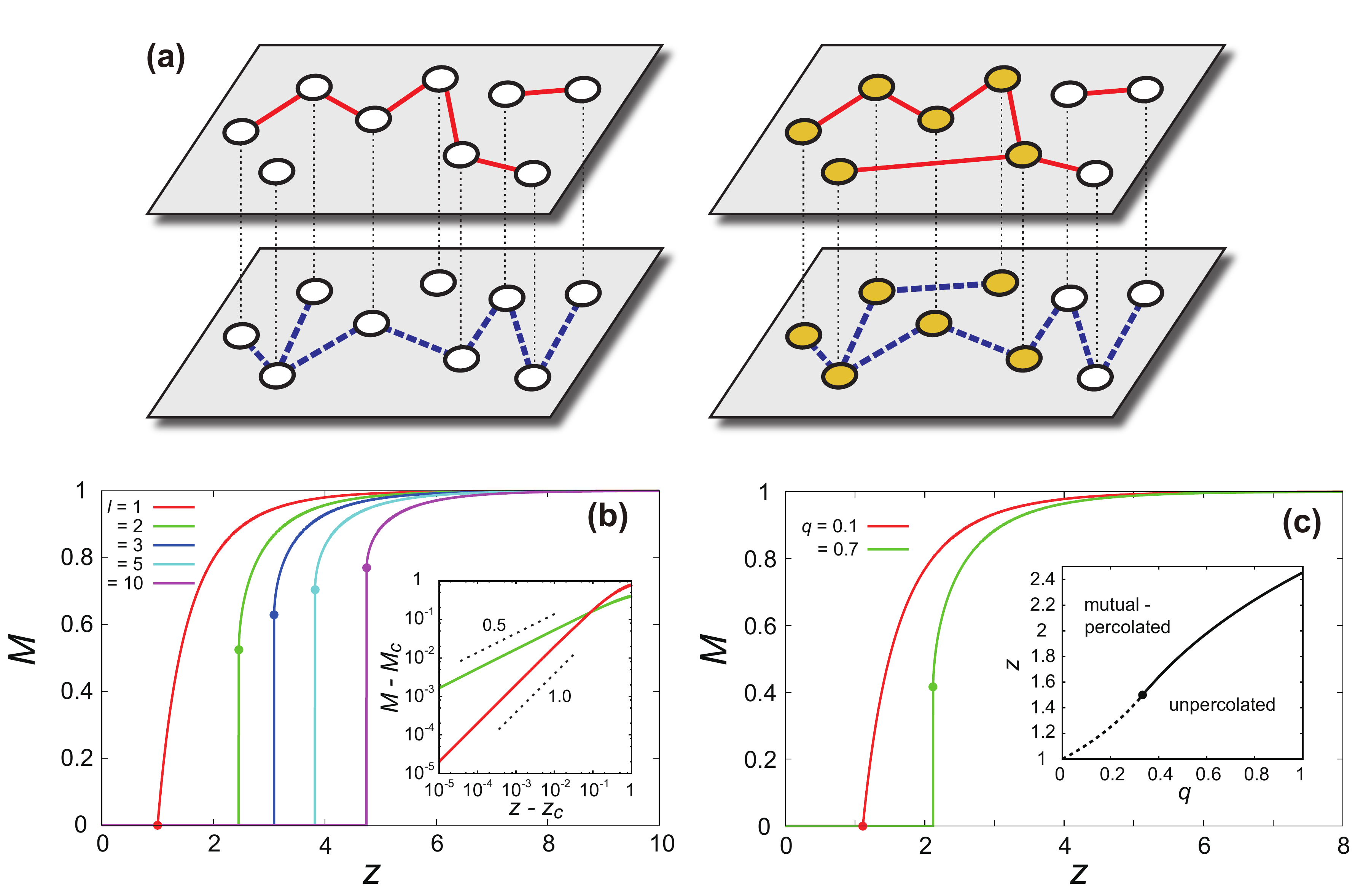}
\caption{(a) Illustration for the notion of mutual percolation. On the left, even though each layer is well connected, the multiplex does not have a mutual component. On the right, the multiplex gains a big mutual component  (marked in orange), after one more link is added to each layer. 
(b) The size of giant mutual component as a function of the layer mean degree $z$ for multiplex ER networks with $\ell=1, 2, 3, 5, 10$ layers. The dot denotes the size of jump at the transition. (Inset) The scaling plot of $M-M_c$ vs.\ $z-z_c$ above the transition, showing the different scaling exponent $\beta=1$ for $\ell=1$ and $\beta=1/2$ for $\ell=2$.
(c) Mutual percolation of the partially-multiplexed ER networks with multiplex fraction $q=0.1$ and $q=0.7$, displaying continuous and discontinuous transitions, respectively. (Inset) The phase diagram in $(q,z)$-plane with mutual-percolated and unpercolated phases separated by lines of continuous (dotted) and discontinuous (solid) transitions, joining at the tricritical-like point (dot).
}
\end{figure*}

\section{Problems with cooperative layer-coupling}
\label{structure}
In many multiplex systems, nodes can be influenced by cooperative activity from multiple layers of the system.
Often the overall synergistic effect is non-additive and nonlinear in individual layer's effect, leading to a situation that conventional single-layer network framework could not address accurately. It therefore has been an outstanding topic in multiplex network studies. In this section we present an overview of the class of problems regarding the structural and dynamical properties of multiplex networks in which the layers are coupled in cooperative manner. 
Problems belonging to this class include (i)  the systems in which the proper functioning of one layer is dependent on the proper function of another layer and {\it vice versa}, often framed in terms of interdependent networks \cite{Buldyrev2010}, and (ii) the processes in which a node's state is affected by synergistic influence from the states of multiple layers, such as multiplex threshold cascade \cite{Brummitt2012,Lee2014}. 
In what follows, we will make concise discussion focusing on how to generalize traditional network theory to multiplex networks with cooperative layers and what the main effects of cooperativity in layer-coupling are, with the examples of mutual percolation, structural robustness, and cascading dynamics.

\subsection{Mutual percolation}
\label{subsec:percolation}
Being connected (or connectivity\footnote{In graph theory, the term connectivity is used to specifically refer to the minimum number of nodes that need to be removed to disconnect the remaining nodes from each other. Here we are using the term in a loose sense.} for short) is a minimal condition for the functionality of a system.
Percolation is a classic statistical physics problem concerning the global connectivity of a system.
One way to generalize the concept of connectivity in multiplex systems with many layers is to require the simultaneous connectivities across different layers, leading to the notions of mutual connectivity and mutual percolation \cite{Buldyrev2010,Son2012}.
Motivations of such definition may come from the interdependent networks of critical infrastructure \cite{Rosato2008} and systems with multiple critical resources \cite{Min2014b}. 
It is a type of cooperative layer-coupling in the sense that each and every layer needs to operate concurrently for the proper functioning of the system as a whole. 

To formalize the problem of mutual percolation, we first define the mutually-connected component (or the mutual component for short) as the set of nodes in which each pair is connected within each and every layer simultaneously \cite{Buldyrev2010,Son2012} (Fig.~4a). 
The mutual component which is extensive in network size, if it exists, is called the giant mutual component. 
The notion of mutual percolation was first introduced in~\cite{Buldyrev2010} while describing the iterative cascade of failures in interdependent networks. 
Key finding of~\cite{Buldyrev2010} was the discontinuous transition in the size of giant mutual component at the critical fraction of random node removals, colloquially referred to as the abrupt collapse of the system. 
Such a discontinuous mutual percolation transition is a strong departure from single-layer network percolations, which is a major reason for much attention from statistical physicists studying network theory \cite{Gao2012a}.

Later the problem was re-casted in purely structural terms in \cite{Son2012},
by noting that at the end of the cascade of failure process all the remaining (unfailed) nodes should form mutually-connected components. 
This formulation allows a more easily-accessible analytical approach \cite{Son2012,Baxter2012} by extending the existing methods from single-layer network theory, which we will take as the basis of the presentation that follows. 
The analytic approach put forward in \cite{Buldyrev2010}, which explicitly tracks the back-and-forth cascade steps of failure, is slightly more involved, yet both approaches yield the equivalent answer when it comes to the giant mutual component.

Mutual components in a given network can be found algorithmically by straightforward iterations towards the requirement of the mutual connectivity~\cite{Buldyrev2010}. More efficient algorithms for finding the mutual components have also been proposed \cite{Schneider2013b,Hwang2014}.

Analytically, the size of giant mutual component of multiplex network can be obtained for networks with locally-treelike undirected uncorrelated layers \cite{Min2014a,Son2012,Baxter2012}, as follows. 
One first consider the probability $w_a$ that the node reached by following a randomly chosen $a$-layer edge does not belong to the giant mutual component. These probabilities satisfy the coupled self-consistency equations,
\begin{align}
1- w_{a} = \sum_{\mathbf{k}}\frac{k_{a}P(\mathbf{k})}{z_{a}}(1-w_{a}^{k_{a}-1})\prod_{b=1,b\neq a}^{\ell}(1-w_{b}^{k_{b}}), 
\end{align}
with $z_a$ is the mean degree of $a$-layer and $a=1,\dots,\ell$.
Terms inside the parentheses give the probability that at least one neighbor in $a$- and all other layers, respectively, of the node belongs to the giant mutual component, the product of which becomes the probability that the node itself belongs to the giant mutual component under the condition of mutual connectivity. 
By averaging this probability over $\mathbf{k}$ weighted by the factor $k_{a}P(\mathbf{k})/z_a$, giving the probability that the multiplex degree of the node reached by following a randomly chosen $a$-layer edge is $\mathbf{k}$, one can end up with equation~(8).
The size of giant mutual component is obtained by the probability $M$ that a randomly chosen node belongs 
to the giant mutual component, which by the same reasoning is given by 
\begin{align}
M = \sum_{\mathbf{k}}P(\mathbf{k})\prod_{a=1}^{\ell}(1-w_{a}^{k_{a}}),
\end{align}
with $w_a$'s being the smallest solution (in the physical range $[0, 1]$) of equation~(8).
Note that this approach reduces to that of the ordinary percolation for single-layer networks for $\ell=1$.

Specific application to multiplex networks with Erd\H{o}s-R\'enyi (ER) layers illustrates the key features.  
For multiplex ER networks with $\ell$ layers of equal mean degree $z$, equations (8) and (9) reduce to a single equation for $M$ as
\begin{align}
M = (1-e^{-zM})^{\ell},
\end{align}
which undergoes a discontinuous transitions from unpercolated $(M=0)$ to mutual-percolated phase $(M>0)$ at the critical layer mean degree, $z_c(\ell)$ (Fig.~4b, main panel). 
For duplex ER networks ($\ell=2$), the transition point is $z_c(2)=2.455407\dots$, which is significantly higher than that of ordinary (single-layer) percolation, $z_c(1)=1$, and the jump in the giant component size at the critical point $M_c(2)=0.511699\dots$ \cite{Buldyrev2010,Son2012}.

The discontinuous mutual percolation transition shows some unusual properties compared to the typical first-order phase transition.  Firstly, the transition does not exhibit hysteresis. Transition occurs at the same point as the mean degree is either increased from below or decreased from above the transition point, providing the basis for the equivalence between the two approaches of references~\cite{Buldyrev2010,Son2012}.
The nature of discontinuous mutual percolation transition was further investigated in reference~\cite{Baxter2012},
showing that the transition is of hybrid type (or mixed order), with discontinuous jump in the order parameter (called the giant viable cluster size in Ref.~\cite{Baxter2012}) followed by the critical scaling, \begin{align}
M-M_c\sim(z-z_c)^{\beta},
\end{align} above the transition point. 
For duplex ER networks, the critical exponent $\beta=1/2$ (Fig.~4b, inset), as in the bootstrap and the $k$-core percolation on single-layer networks \cite{Baxter2010,Baxter2011}.
The average size of finite mutual components, an analogous quantity to the susceptibility in ordinary percolation, does not diverge at the transition point. 
Instead, the critical scaling above the transition point is attributed to the diverging scale of avalanches upon the node removal as the transition point is approached from above~\cite{Baxter2012}. 

\subsubsection{Partial multiplexity}
Situation may occur that some of the nodes in a multiplex do not participate in all the layers, a case that might be called the partial multiplexity. This generalization has been studied in both approaches~\cite{Son2012,Parshani2010a}. Let us consider the multiplex network in which the fraction $q$ of nodes participate in both layers (multiplex nodes) but the remaining $1-q$ fraction of nodes in each layer are specific to that layer. 
The generalized mutual component is defined as the set of connected nodes in which every multiplex node is connected in both layers simultaneously. In general the size of generalized mutual component in different layers can be different. For duplex ER networks, the sizes of generalized mutual components in the two layers are obtained by generalizing equation~(10) with $\ell=2$. It reads~\cite{Son2012}
\begin{align} 
M_A &= (1-e^{-z_AM_A})(1-qe^{-z_BM_B}),  \nonumber\\
M_B &= (1-qe^{-z_AM_A})(1-e^{-z_BM_B}), 
\end{align}
where $M_a$ and $z_a$ are the generalized mutual component size and the mean degree of $a$-layer. One can see that $M_A\neq M_B$ when $z_A\neq z_B$ in general with $q\neq1$. A key feature of this model is that depending on $q$ and $z_i$'s the nature of transition can change from discontinuous to continuous, exhibiting the tricritical-like point \cite{Son2012} (Fig.~4c).

In the framework of interdependent networks, more general scenario was considered \cite{Son2012,Parshani2010a}, such that in $A$-layer a fraction $q_A$ of nodes are dependent on a node in $B$-layer (and vice versa). The dependency relation can also be unidirectional, making it more general than the simple multiplex networks. 
It can be treated analytically in similar way and gives qualitatively similar picture as the previous partial multiplex case. Through this setting with $z_A=z_B$, the critical behavior of the giant generalized mutual component was obtained, as $M_A\sim (q_A-q_{A}^*)^{\beta}$ with $\beta=1$ along the line of discontinuous transitions whereas $\beta=1/2$ when $q_B=q_{B}^*$ is fixed, with $(q_{A}^*,q_{B}^*)$ denoting the point at which the lines of discontinuous and continuous transitions join \cite{Parshani2010a}. A phenomenological analogy to liquid-gas transition exhibiting the critical point was proposed in \cite{Parshani2010a}, but further understanding of the nature of symmetry-breaking in this problem is required to establish a more complete correspondence.
Generalizations of partially interdependent networks with both dependency and connectivity links have been studied in \cite{Parshani2011,Hu2011}.

\subsubsection{Network of networks} 
In the framework of interdependent layers, one can further extend to the general case of the network of  interdependent layers, dubbed as ``the network of networks'' (NoN) and  studied for a number of different topology \cite{Gao2011,Gao2012b,Gao2013}. 
A notable result for NoN is that when the interdependency between networks (layers) is such that there is a one-to-one dependency relation between all the nodes in each pair of connected layers, the mutual component does not depend on the topology of NoN \cite{Bianconi2014a}. It turns out that the mutual component of such NoN coincides with that of the fully-connected NoN constructed with the same set of layers, {\it i.e.}, the simple multiplex network.
As an instance in which the interdependency between layers is not simple, the so-called configuration model of  NoN was studied~\cite{Bianconi2014b}. In this model the nodes are assigned  a given number $Q_a$ of interlayer links in each layer $a$, which are connected in the configuration-type algorithm. A peculiar behavior is found by using the message passing calculation, that the NoN can undergo a series of multiple transitions when the interlayer degrees $Q_a$ are distributed heterogeneously~\cite{Bianconi2014b}. 

\subsubsection{Spatially embedded multiplexes} 

Many complex systems existing in the real world are embedded in two dimensional (2D) Euclidean space to some extent~\cite{Barthelemy2011}, as in the cases of multiplex transportation system or interdependent critical infrastructure.
The mutual percolation problem for spatially embedded multiplexes poses theoretical challenge as well to cope with the correlation imposed by the spatial constraints. 
First, the mutual percolation of duplexes with diluted 2D square lattice was studied \cite{Son2011}, claiming that the mutual percolation transition becomes continuous and the order parameter exponent $\beta$ could be larger than that of ordinary percolation, suggesting that the mutual percolation transition can be less abrupt than ordinary percolation transition, in sharp contrast to the cases of random networks. Specifically in 2D, it was estimated that $\beta=0.171(2)$ for duplex square lattices, compared to $\beta_{\rm single}=5/36=0.1386\dots$ for a single-layer 2D square lattice. This claim was challenged in reference~\cite{Berezin2013b}, arguing that the exponent $\beta$ should have the same value in both duplex and single-layer lattices. Lacking the exact solution for the multiplex lattices, however, the conflict has not been resolved completely as yet.

Later, a variant of multiplex lattice model, called the interdependent lattice networks, in which the node in one lattice makes a dependency link to a node in the other lattice randomly chosen within a certain distance $r$ \cite{Li2012}. For $r=0$ the model reduces to the multiplex lattice model of~\cite{Son2011,Berezin2013b}. As $r\to\infty$, the two lattices are coupled by completely random dependency links, thereby one can expect the transition to be discontinuous as in the multiplex random networks. In between there might exist a point $r^*$ separating the two behaviors. For two-layer interdependent lattices it was estimated that $r^*\approx8$ (in lattice unit) \cite{Li2012}. For $r<r^*$, the transition is continuous, whereas it becomes discontinuous for $r>r^*$.  
More recently, it was shown that for $r\to\infty$, the mutual percolation transition in even partially-interdependent lattices is always discontinuous as long as the fraction of dependent nodes $q>0$, 
that is, any nonzero fraction of dependent nodes can drive the transition to be discontinuous \cite{Bashan2013}.
This result was interpreted that the spatially embedded multiplex networks can be extremely more broadly vulnerable to abrupt (discontinuous) collapse than random networks. 

\begin{figure}
\centering
\includegraphics[width= .8\linewidth]{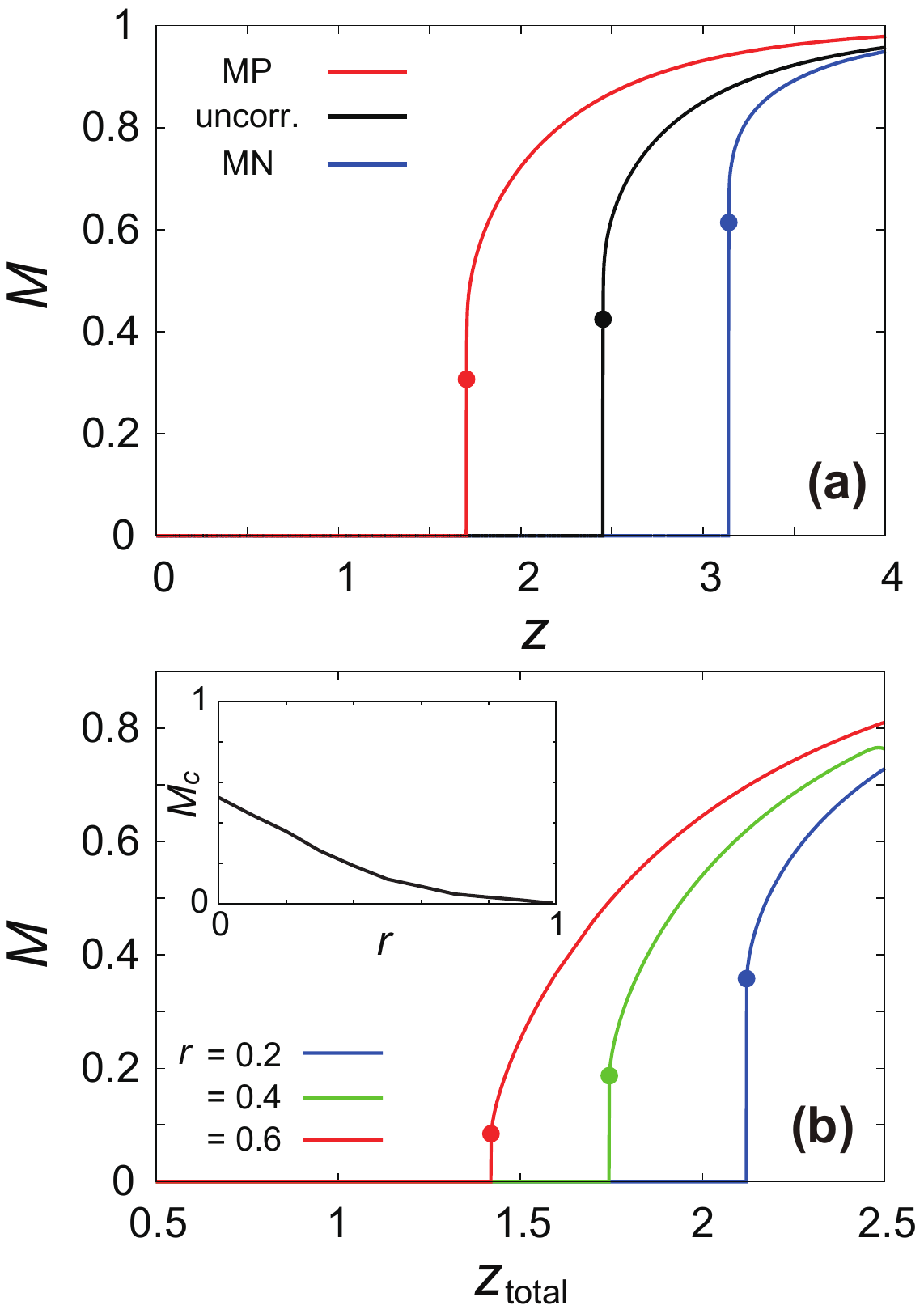}
\caption{Mutual percolation of duplex ER networks with (a) interlayer degree correlation of three different kinds, the MP, uncorrelated, and MN couplings, and (b) link overlap with different overlap fraction $r=0.2, 0.4, 0.6$. (Inset) The size of jump $M_c$ at the transition as a function of the overlap fraction $r$, which decreases gradually to zero as $r\to1$. Adapted from references~\cite{Min2014a} (a) and \cite{Min2014c} (b).
}
\end{figure}

\subsubsection{Effect of correlated multiplexity}  
The effect of correlated multiplexity on mutual percolation has been addressed in terms of the interlayer degree correlations \cite{Parshani2010,Min2014a,Buldyrev2011,Valdez2013} and the link overlap \cite{Hu2013,Cellai2013,Min2014c}. 
The most commonly-observed interlayer degree correlation would be a positive interlayer degree correlation, the effect of which has been the major subject of studies~\cite{Parshani2010,Buldyrev2011,Valdez2013}. 
The main result of those studies is that the positive interlayer degree correlation (also termed the intersimilarity in Ref.~\cite{Parshani2010}) makes the multiplex system more robust to random failure \cite{Parshani2010,Min2014a,Buldyrev2011}, in the sense that the mutual percolation transition point becomes lower  (Fig.~5a). The study has been extended to include negative interlayer degree correlation as well~\cite{Min2014a}. 
More recently, in the framework of interconnected networks a particular form of positively correlated interlayer connection pattern was introduced as an optimal coupling architecture for robustness of the mutual connectivity to random failure~\cite{Reis2014}, offering a consistent perspective with the earlier findings from the multiplex and interdependent networks.  It was also proposed that the suggested optimal pattern might underlie the topological stability of the interconnected functional brain networks \cite{Reis2014}. To what extent such broad perspective of topological stability could extend to the actual functional stability of given multiplex systems remains an outstanding open problem.

Another pervasive kind of correlated multiplexity is the link overlap. The overlapping link plays a distinctive role in mutual percolation, in that the nodes connected by overlapping links form a mutual component by themselves. Therefore the link overlap would facilitate mutual percolation. In the extreme case where every link is overlapping link, all the layers becomes identical and the mutual percolation problem reduces to ordinary percolation of the single layer, showing continuous transition. A primary question of theoretical interest would then be the existence of the nontrivial value of the fraction of overlapping links $r$ across which the nature of transition changes from discontinuous (as in $r=0$) to continuous (as in $r=1$). This question was examined by two groups independently using different methodology \cite{Hu2013,Cellai2013}, only to reach conflicting conclusions. The two approaches were later scrutinized in \cite{Min2014c}, to show that they in general apply to different percolation processes (see the next section for more details). 

To illustrate the effect of link overlap, let us consider the case of duplex ER networks of layers with equal mean degrees $z_{\rm total}$ among which the fraction $r$ is the overlapping link. As the overlap cluster forms a mutual component by itself, it is useful to renormalize it as a supernode, which is connected by non-overlapping links to other supernodes (overlap clusters). To obtain the giant mutual component size, one needs to augment equation~(10) by taking into account the sizes of supernodes, the distribution of which is given by the component size distribution of the network formed by overlapping links, denoted $R(m)$. The giant mutual component size $M$ is then given by \cite{Hu2013,Min2014c}
\begin{align}
M = R_{\infty} + \sum_{m=1}^{\infty} R(m) (1-e^{-mzM})^{2},
\end{align}
where $R_{\infty}$ is the probability that a randomly chosen node belongs to the infinite-size overlap cluster  and $z$ is the mean degree for non-overlapping links, $z=z_{\rm total}(1-r)$. Solving this, it was obtained that the transition remains discontinuous as long as the overlap fraction $r<1$ (Fig.~5b), with the jump $M_c$ at the transition vanishes gradually as $r\to1$ (Fig.~5b, inset); the tricritical point claimed in \cite{Cellai2013} is found absent. This means that despite its facilitating role, the link overlap {\it per se} is insufficient to change the nature of transition in multiplex ER networks. Yet the possibility of nontrivial tricritical point in layers of other topology remains open. Furthermore, it is left as a theoretical challenge to extend the analytic theory beyond the simple multiplex cases \cite{Min2014c}.

\subsubsection{Related models: Viability, weak and $\mathbf{k}$-core percolation}
\label{subsec:viability}
In the mutual percolation problem, an implicit assumption was that the connectivity by itself is sufficient for functionality of the system. In some multiplex systems, the connectedness among nodes {\it per se} may not be sufficient but the connected component can be functional only when it is connected also to the resources essential for the functionality of the system. A representative example in this class is the modern city, which is supported by multiple layers of resource supply network such as power and water supply networks. These vital resources are not produced in every node; rather they are generated from a specific set of nodes (power stations and water sources, respectively). Taking this feature into account, a model of viability of multiplex system with multiple resource demands was proposed \cite{Min2014b}, which is defined as follows. 

Consider a network with $\ell$-multiple layers, where each layer of the multiplex network corresponds to a certain resource supply network. A given fraction $\rho$ of nodes (called the resource nodes) generates and distributes resources essential to be viable. Only viable nodes can function properly and transmit resources further to their connected neighbors. Then, a node is viable only if it can reach, via other viable nodes, to a resource node in each and every layer. The viability $V$ of the system is defined as the fraction of viable nodes, given the network parameters and distribution of resource nodes. To compute the viability $V$, one can set up equations analogous to those for the mutual percolation, equations~(8) and (9) as~\cite{Min2014b}
\begin{align}
1-v_{a} &= \rho+(1-\rho)\sum_{\mathbf{k}}\frac{k_{a}P(\mathbf{k})}{z_{a}}(1-v_{a}^{k_{a}-1})\nonumber\\
&\times\prod_{\substack{b=1 \\b\ne a}}^{\ell}(1-v_{b}^{k_{b}}), 
\end{align}
where $v_a$ is the probability that the node reached by following a randomly chosen $a$-type link $(a = 1,...,\ell)$ is not viable, and 
\begin{align}
V = \rho+(1-\rho)\sum_{\mathbf{k}}P(\mathbf{k})\prod_{a=1}^{\ell}(1-v_{a}^{k_{a}}).
\end{align}
In equations~(14) and (15), the first term on the right hand side is the probability that the chosen node is a resource node and the second term gives the probability that it is not a resource node but is connected to the giant viable cluster.

\begin{figure}
\label{fig:overlap-viability}
\centering
\includegraphics[width= .9\linewidth]{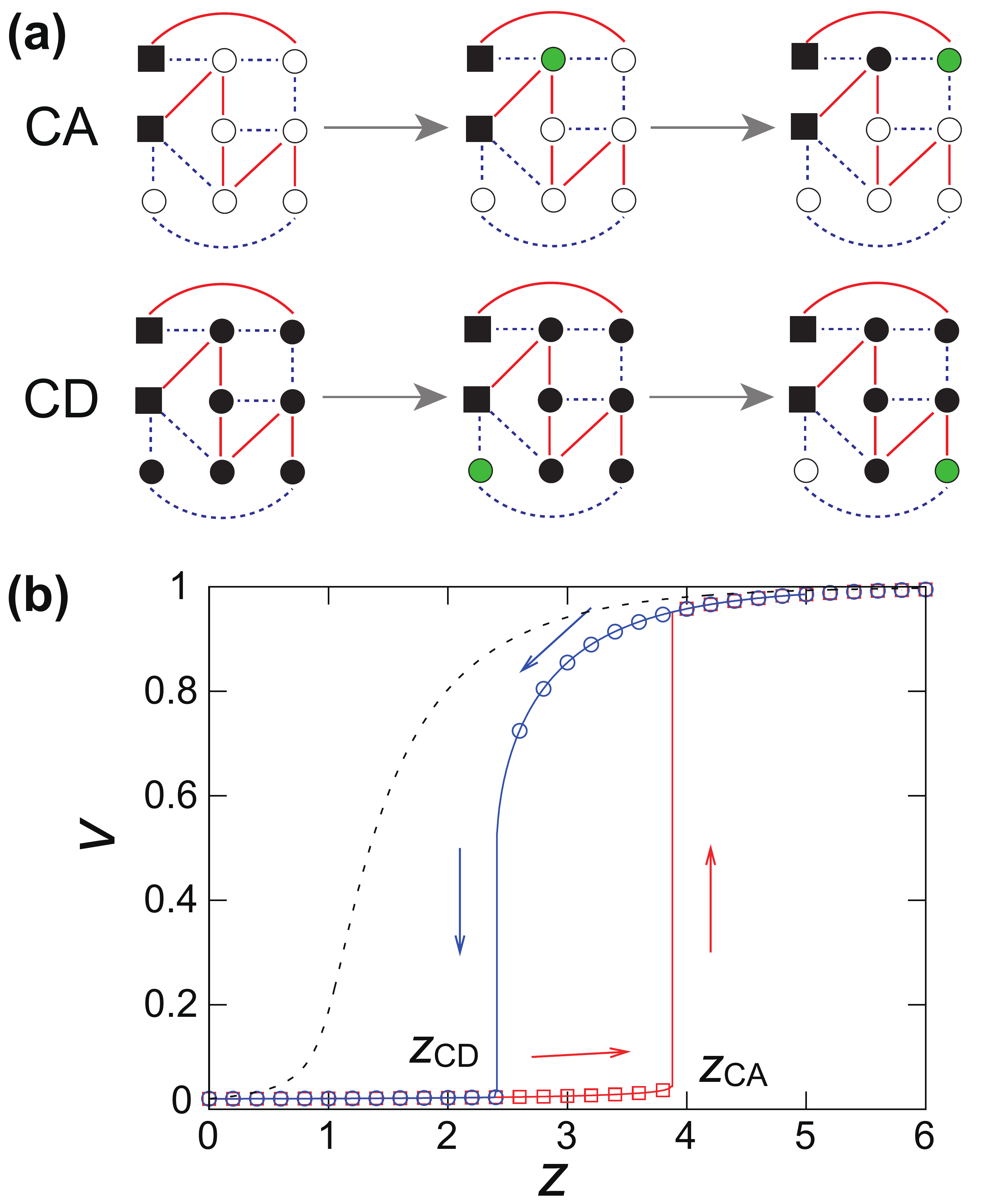}
\caption{(a) Illustrations of the CA and the CD algorithms for viability. Resource nodes (squares) generate resources. If a node connects with resource nodes through each type of link, denoted by the solid and dashed lines, the node is viable (filled circles) and can transmit resources further to its neighbors. If not, the node is unviable (open circles). Shaded (green) circles denote the node whose state is to be updated (activated in CA and deactivated in CD) at each step. 
(b) Hysteresis curve of $V$ with $\rho_0=0.02$. Starting from the well-connected high viability state, the systemic collapse $(\circ)$ and the subsequent recovery $(\Box)$ exhibit different curves. Dashed line indicates $\ell=1$ case for comparison, without hysteresis. Adapted from reference~\cite{Min2014b}.
}
\end{figure}

Applying to duplex ER networks with equal layer mean degree $z$, one has the single equation for $V$ analogous to equation~(10) as 
\begin{align}
V =\rho+(1-\rho)(1-e^{-zV})^2.
\end{align}
Solving equation~(16), one obtains two stable solutions for a range of parameters $\rho$ and $z$, suggesting the bistability in viability. 
Viability undergoes discontinuous jump as one of the two stable solutions loses its stability. The two solutions are shown to correspond to outcomes of different algorithms for identifying the viable components, called the cascade of activations (CA) and the cascade of deactivations (CD), respectively \cite{Min2014b} (Fig.~6a). This means that the viability of a system depends also on the way it reached the current state, that is, the viability exhibits hysteresis (Fig.~6b).
In the limit $\rho\to0$, equations~(14)--(16) reduce to those for mutual percolation equations~(8)--(10). In this sense the viability model can be considered as a generalization of mutual percolation. In this limit, only the CD branch remains meaningful (the CA branch solution becomes trivial, $V=0$ for all $z$), which coincides with the solution for mutual percolation. 

In the presence of link overlap, richer picture emerges \cite{Min2014c}. 
Viability still exhibits bistability, corresponding to the outcomes of the CA and CD algorithms, respectively, but they do not come from the multiple solutions of a single equation but are obtained as the solutions of two different equations describing the CA and CD algorithms, respectively. Interestingly, the two different approaches proposed for solving the mutual percolation problem with link overlap~\cite{Hu2013,Cellai2013} describe the two algorithms, the CD algorithm by the method of~\cite{Hu2013} and the CA by that of~\cite{Cellai2013}. This finding suggests that  the multiplex system can in general respond to activation and deactivation processes in inherently different ways. 

This property has also been observed in similar percolation models called the weak bootstrap and pruning percolations~\cite{Baxter2014}. 
The weak bootstrap percolation proposed in~\cite{Baxter2014} is the same as the CA algorithm in the viability problem, whereas the weak pruning percolation is similar but not identical to the CD algorithm\footnote{It might be worthwhile to note that this nomenclature is somewhat unfortunate, in that the original bootstrap percolation \cite{Chalupa1979} is defined as a pruning process, and the corresponding activating process is known as diffusion percolation \cite{Adler1988}.}, and 
it is found that the two percolation processes yield different results for multiplexes \cite{Baxter2014}. 
$k$-core percolation was generalized to multiplex networks into $\bf k$-core percolation~\cite{Azimi-Tafreshi2014a}, in which {\bf k}-core is defined as the largest subgraph with each node therein having at least $k_{a}$ edges within the subgraph in each layer, with ${\mathbf k}=\{k_a\}$. Hybrid phase transition akin to that of mutual percolation was found except for $(k_{A},k_{B})=(1,1)$, for which the transition is continuous~\cite{Azimi-Tafreshi2014a}. 

\subsection{Robustness against attacks}
\label{subsec:robustness}
Network robustness has been a major topic of network theory \cite{HavlinBook}, formulated as various percolation problems. Compared to the usual percolation problem concerning random dilution \cite{Cohen2000}, the robustness problem focuses on the response of the system to a variety of different ways of dilution, such as node removals according to their degrees, commonly referred to as the intentional attack~\cite{Albert2000,Cohen2001}. 

For the class of multiplex networks with cooperative layer coupling, it is the giant mutual component that is the primary quantity for addressing the system's robustness. The size of the giant mutual component $M$ of multiplex networks with locally-treelike uncorrelated layers after the node removals can be obtained as the generalization of equations~(8) and (9) by following reference~\cite{Callaway2000}. It reads \cite{Min2014a}
\begin{align}
1-w_{a} &= \sum_{\mathbf{k}}\frac{k_{a}P(\mathbf{k})}{z_{a}}[1-\phi(\mathbf{k})]\nonumber\\
&\times(1-w_{a}^{k_{a}-1})\prod_{\substack{b=1 \\ b\ne a}}^{\ell}(1-w_{b}^{k_{b}}), \\
M &= \sum_{\mathbf{k}}P(\mathbf{k})[1-\phi(\mathbf{k})]\prod_{a=1}^{\ell}(1-w_{a}^{k_{a}}),
\end{align}
where the function $\phi(\mathbf{k})$ encodes the removal strategy in terms of the probability that the node with multiplex degree $\mathbf{k}$ is removed.  For instance, for the uniform random removal of fraction $f$ of nodes (as in usual site percolation), $\phi=f$ (constant). For the intentional attack based on the node's total degree $K=\sum_{a=1}^{\ell}k_{a}$, $\phi={\mathrm\Theta}(K-K_{c})$, where  $\mathrm\Theta$ denotes the Heaviside step function and $K_{c}$ is the cutoff total degree for the attack~\cite{Min2014a}. 
Several attack strategies and multiplex scenarios have been studied such as the targeted attack with $\phi(k)\propto k^{\alpha}$ with adjustable parameter $\alpha$ on fully multiplexed networks (in the form of interdependent networks) \cite{Huang2011}, on partial multiplexes \cite{Dong2012}, and on network of networks \cite{Dong2013b}. Localized attack on spatially embedded networks was also studied \cite{Berezin2013}. 

Network robustness property would also be strongly influenced by the correlated multiplexity \cite{Min2014a,Min2014c}. Against random failures, the positive (negative) interlayer degree correlation is shown to enhance (decrease) the robustness of mutual connectivity in multiplex random networks. For the case of total-degree based attack, the vulnerability depends not only on the correlated couplings but also on the initial density of the networks~\cite{Min2014a}. Such non-monotonic and protocol specific responses of multiplex networks to damage could complicate the design of robust coupled systems \cite{Reis2014,Yagan2012,Schneider2013}, as the optimal structure could strongly depend on the type of damage considered. 

\subsection{Cascades and complex contagion}
\label{subsec:cascade}

Percolation problems are utterly concerned with the structural or topological connectivity of the system. Although the network structure itself can be a strong determinant of the system's function and robustness, there are many other phenomena in which the specific dynamical process occurring on top of the structure also confers important consequences~\cite{BarratBook}. In such phenomena, besides the network structure, each node is endowed some attribute index (or state function) which is affected by the neighbors' states. The particular way the nodes' states affect one another defines the dynamical processes. In this section we survey a particular class of dynamical processes for which the cooperative layer coupling has been a crucial ingredient when generalized to multiplex systems, the cascade models. It includes the Watts-type threshold cascade models \cite{Brummitt2012,Lee2014,Yagan2012b}, the sandpile models \cite{Lee2012b,Brummitt2012b}, and the Motter-Lai-type overload cascade models \cite{Tan2013}. 

The threshold cascade model began as a model of how the behavioral adoption spreads over the social community \cite{Schelling1973}, and was later generalized and formulated further by \cite{Granovetter1978,Watts2002}. A simple mathematical formulation by Duncan Watts \cite{Watts2002} has become a standard model, called accordingly the Watts model. Each individual (node) can in one of two states, active and inactive. An inactive node decides to activate if the fraction of neighbors who are already active exceeds the threshold. This activation process occurs as a cascade until no further activation can be made. It is known that in networks, depending on network parameters and threshold distribution the so-called global cascade might occur, by which a finite fraction of large (infinite, theoretically) network can be activated from a vanishingly small fraction of initially-active seed nodes \cite{Watts2002}. 

This model has been extended to multiplex networks \cite{Brummitt2012,Lee2014}, in which the social influences from different layers are integrated non-additively and nonlinearly. 
The analytical approach applicable to locally-treelike uncorrelated layers proceeds in a similar but slightly different logic as that of mutual percolation (see Refs.~\cite{Brummitt2012,Lee2014} for detailed arguments). 
The cascade size $\rho$ from random-distributed initial seeds of fraction $\rho_{0}$ can be calculated as
\begin{align}
\rho = \rho_{0} + (1-\rho_{0}) \sum_{\mathbf{k}}P(\mathbf{k})\sum_{\bf{m}=0}^{\mathbf{k}}\prod_{a=1}^{\ell}B_{m_{a}}^{k_{a}}(q_{\infty}^{(a)})\bar{F}(\mathbf{m},\mathbf{k}),
\end{align}
where $B_{m}^{k}(q)$ is the shorthand notation for binomial distribution, ${k \choose m} q^{m}(1-q)^{k-m}$. The quantity $\{q_{\infty}^{(a)}\}$ is the fixed point of the coupled recursion relation,
\begin{align}
q_{n+1}^{(a)} &= \rho_{0} + (1-\rho_{0})\sum_{\mathbf{k}}\frac{k_{a}P(\mathbf{k})}{z_{a}}\nonumber\\
&\times\sum_{m_{a}=0}^{k_{a}-1}\sum_{\{m_{b}\}=\mathbf{0}, b\neq a}^{\{k_{b}\}}B_{m_{a}}^{k_{a}-1}(q_{n}^{(a)})\nonumber\\
&\times\prod_{b\neq a}B_{m_{b}}^{k_{b}}(q_{n}^{(b)})\bar{F}(\mathbf{m},\mathbf{k}),
\end{align}
starting from every $q_0^{(a)}=\rho_0$.
$\bar{F}(\mathbf{m},\mathbf{k})$ is the response function encoding the activation rule, giving the probability that the node with $\mathbf{m}=\{m_a\}$ active neighbors among its multiplex degree $\mathbf{k}$ will activate. 
This analytical approach could further be generalized to multiplexes with correlated and modular layers by employing methods devised for the single-layer networks \cite{Gleeson2008,Melnik2014}.

\begin{figure}
\centering
\includegraphics[width= .77\linewidth]{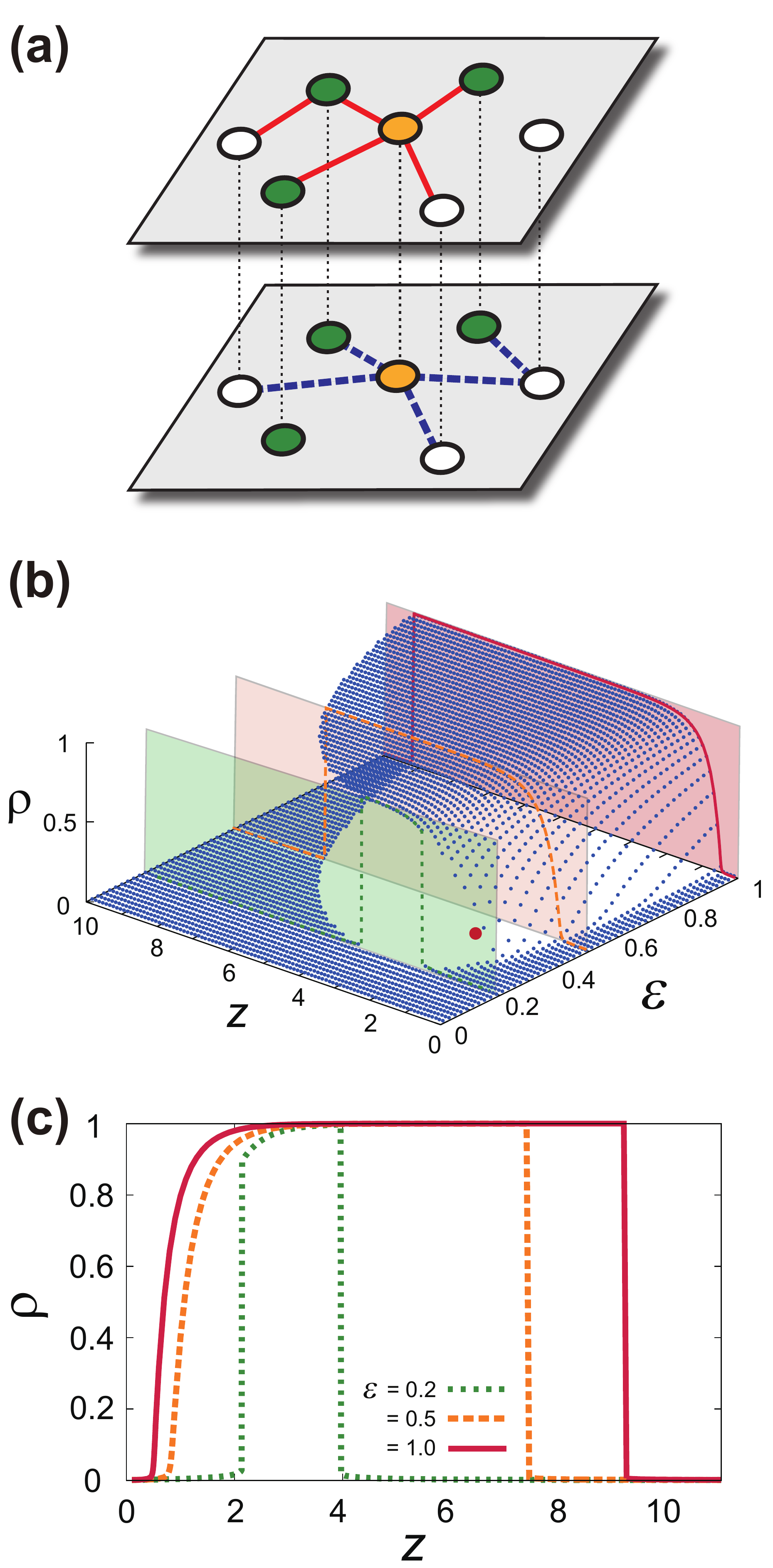}
\caption{(a) An example multiplex configuration for illustrating the two response rules. Assuming the uniform threshold for activation $R=1/2$, green (dark) circles denote active nodes and the light (white, orange) ones the inactive nodes. The orange node will activate itself under the OR rule, but will remain inactive under the AND rule.
(b) The cascade size $\rho$ with the initial seed fraction $\rho_0=0.001$ and the uniform threshold $R=0.18$ as a function of the layer mean degree, $z$, and the fraction of OR nodes, $\mathcal{E}$. Red dot marks the cusp point at which the lines of continuous and discontinuous transition join. Three vertical planes with the guideline denote the slices for which the panel (c) is drawn. 
(c) The cascade size $\rho$, with the same parameter as (b), as a function of $z$, for three different $\mathcal{E}=0.2$  (green dotted), $0.5$ (orange dashed), and $1.0$ (red solid). 
For $\mathcal{E}=0.2$ the transition to global cascade occur discontinuously, whereas it is continuous for the other two cases. Adapted from reference~\cite{Lee2014}.
}
\vspace{-.3cm}
\end{figure}

Two simple yet representative activation rules are considered (Fig.~7a): (i) the OR rule, in which the node activates once the threshold is met in at least one layer, and (ii) the AND rule, in which the node activates only after the threshold is met in all its social layers. Intuitively, the OR rule would facilitate the cascade, whereas the AND rule will impede it, which are confirmed by both theory and simulations \cite{Brummitt2012,Lee2014}. A less intuitively clear picture emerges when the two activation rules are mixed in population \cite{Lee2014}. As the fraction of nodes following the AND rule increases, there exists a critical fraction above which the transition to global cascade becomes discontinuous, compared to the continuous transition below this fraction (Figs.~7b and 7c), followed by tricritical-like scaling behavior \cite{Lee2014}.  
This is another instance in which the cooperative layer coupling, now driven by dynamics, induces discontinuous transition in multiplex systems. Different multiplex generalization of Watts model has also been studied \cite{Yagan2012b}. 

Another prototypical model of cascades on networks is the model of cascading failure based on loads, put forward by Motter and Lai~\cite{Motter2002}. 
Each node $i$ is assigned its own capacity, $C_i=(1+\alpha)L_i$, 
setting the maximum load that the node can endure, where $\alpha$ is the tolerance parameter and $L_i$ is the load defined as the total number of shortest paths passing through that node $i$~\cite{Goh2001}. Initiated by the failure (removal) of a small set of nodes, the global redistribution of the shortest paths and thereby the load can induce the cascade of failure of overloaded nodes whose load becomes to exceed the capacity. 
Its distinctive property is that the cascade occurs in a nonlocal manner, unlike the threshold cascade models introduced above. The Motter-Lai-type models have been studied on multiplex networks with a particular focus on the effect of interlayer degree correlation and the optimal coupling patterns~\cite{Tan2013}. 
For instance, it has shown that the positive interlayer correlation is beneficial in mitigating the cascades compared with other coupling patterns, in line with the findings in mutual percolation \cite{Tan2013}.

\section{Problems with complementary layer-coupling}
\label{dynamics}

Different layers in the coupled transportation system provide alternative or complementary means of traveling from one place to another. In such systems, malfunction of one single layer does not fundamentally alter the functionality of the whole system. In this section we consider the problems arising in such multiplex systems with complementary layers. Complementary layers are coupled in connective way\footnote{In this respect many problems in this class can also be formulated as the interconnected networks, especially in terms of mathematics.}, and often the whole system can be regarded as a single super-network in which the network layers connected by interlayer links. Hence the problems in this class are admittedly much more straightforward to generalize from single-layer framework than the problems with, say, cooperative layers. 
This aspect might have played a role in prompting skepticism around the whole multiplex network framework.
Notwithstanding, it is still useful to retain the multiplex framework in such problems for the sake of generality and methodological clarity, as we shall argue below.
 
\subsection{Percolation}

Percolation of a system of $\ell$ interconnected layers was first addressed systematically in~\cite{Leicht2009}\footnote{Comparison of the two seminal papers \cite{Leicht2009,Buldyrev2010} may reveal an interesting perspective regarding how the community perceives on multiplex problems of different kinds. Both works were, to our knowledge, first presented to wide public at the International Conference on Network Science (NETSCI) in May 2009 and uploaded in parallel onto arXiv that summer. Buldyrev {\it et al.}'s work on mutual percolation was eventually published in Nature in 2010~\cite{Buldyrev2010} and is by now cited about 800 times according to Google Scholar; Leicht and D'Souza's work is still unpublished \cite{Leicht2009}, with some 50 citations thus~far.},  although there had been related  predecessors like the graphs with colored edges \cite{Soderberg2003a,Soderberg2003b}. 
In this problem, one is primarily interested the existence of the giant component, that is the extensive subset of nodes which are connected through links regardless of their types (layers). They developed a generating function method, by generalizing the well-established single-layer framework for the interconnected layers \cite{Leicht2009}. Methodologically this framework could also find predecessors in the percolation of clustered single-layer networks~\cite{Newman2009}. 
Below we will present the framework applied specifically to simple multiplexes \cite{KMLeeBook,Min2014a}. The general framework is not too different, and the interested readers are referred to \cite{Leicht2009}. 

Given the multiplex degree distribution, $P(\mathbf{k})$, and its generating function,
\begin{align}
G_{0}(\mathbf{x})= \sum_{\mathbf{k}}P(\mathbf{k})\prod_{a=1}^{\ell}x_{a}^{k_{a}}
\end{align}
with $\mathbf{x}=\{x_a\}$,
the generating function for the remaining degree distribution after following a randomly chosen $a$-layer edge is obtained as 
\begin{align}
G_{1}^{(a)}(\mathbf{x}) = \frac{1}{z_{a}}\frac{\partial}{\partial x_{a}}G_{0}(\mathbf{x}).
\end{align}
The giant component size $S$ of the multiplex of locally treelike uncorrelated layers can be obtained using these generating functions as \cite{KMLeeBook,Min2014a} 
\begin{align}
S = 1 - G_{0}(\mathbf{u}),
\end{align}
where $\mathbf{u}=\{u_a\}$ is the probability that a node reached by following a randomly chosen $a$-layer edge does not belong to the giant component, satisfying the coupled self-consistency equations,
\begin{align}
u_{a} = G_{1}^{(a)}(\mathbf{u}), \qquad \textrm{with } a = 1,2,\dots,\ell.
\end{align}
The size $B$ of the giant bicomponent defined as the extensive subset of nodes connected by at least two disjoint paths \cite{Newman2008} is also calculated for multiplex networks as~\cite{Min2014a}
\begin{align}
B = 1 - G_{0}(\mathbf{u}) - \sum_{a}(1-u_{a})z_{a}G^{(a)}_{1}(\mathbf{u}).
\end{align}
The last term gives the difference between the giant component $S$ and bicomponent $B$. The condition for existence of the giant component $S > 0$ and that for the giant bicomponent $B > 0$ coincide and is given as that the largest eigenvalue of the Jacobian matrix $\mathbf{J}$ of equation~(24) at the trivial solution $\mathbf{u}=\{1,\dots,1\}$ be larger than unity~\cite{Min2014a}. 

\begin{figure}
\centering
\includegraphics[width= .9\linewidth]{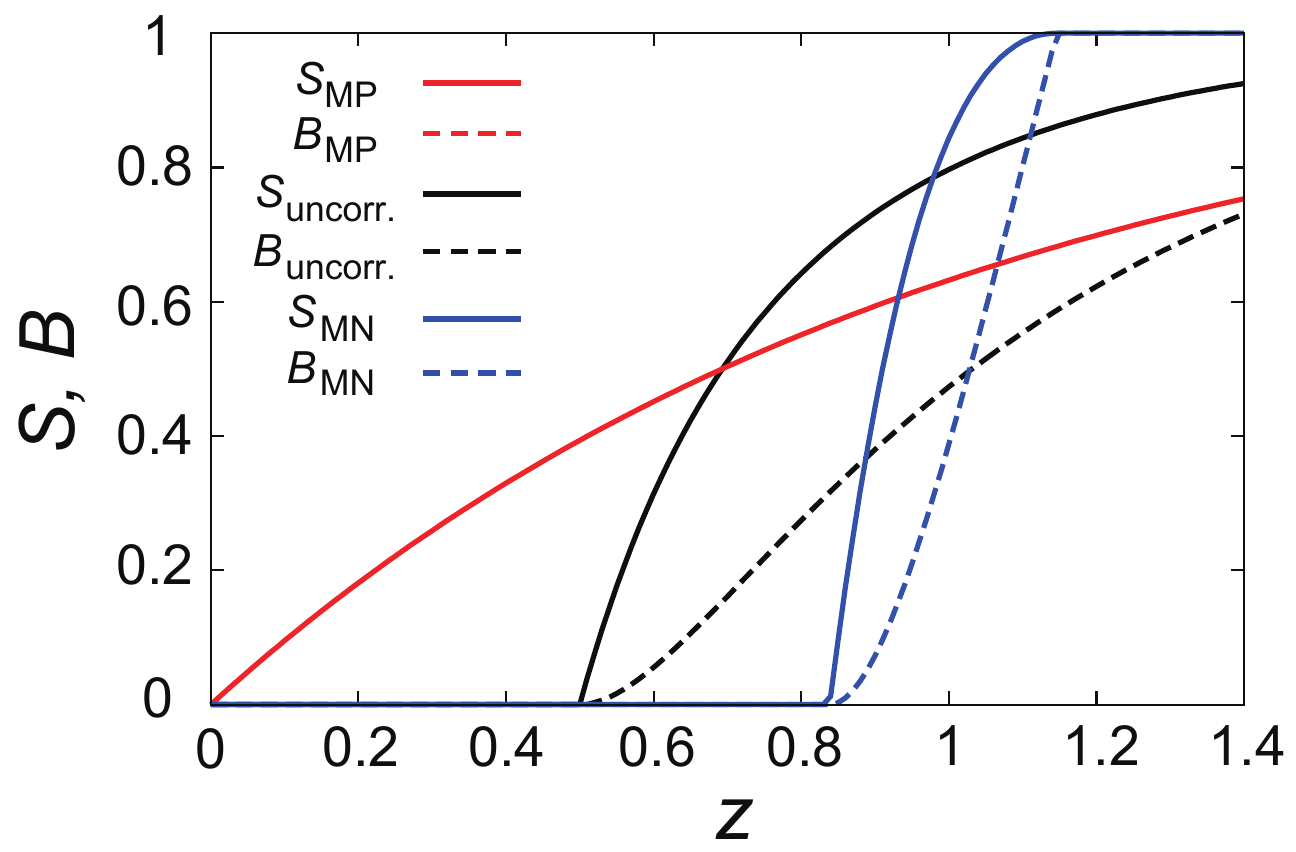}
\caption{Sizes of the giant component $S$ (solid lines) and the giant bicomponent $B$ (dotted lines) for the MP (red), the uncorrelated (black), and the MN (blue) couplings of duplex ER networks. Note that the two curves for the MP coupling overlap completely. Adapted from reference~\cite{Min2014a}.}
\end{figure}

Applying to duplex ER networks with layers of equal mean degree $z$, a number of peculiar behaviors of correlated multiplex random networks have been observed (Fig.~8). First, the MP interlayer degree correlation was shown to promote the percolation to the extreme, in that the percolation threshold decreases all the way to $z_c=0$. On the other hand, in the MN correlated duplexes, the percolation is significantly delayed as $z_c=0.838587\dots$, compared to $z_c=1/2$ for the uncorrelated coupling, but once it occurs the giant component grows more quickly and can span the whole network $(S=1)$ at a finite mean degree $z^*=1.146193\dots$ per layer \cite{KMLee2012,KMLeeBook}.
As for the giant bicomponent, in MP correlated duplexes its size $B$ is obtained identical with the giant component size $S$, meaning the giant component itself becomes a bicomponent, whereas in general the giant bicomponent grows more slowly than the giant component~\cite{Min2014a}.

\subsection{Diffusion}
\label{subsec:diffusion}

Diffusion is one of the simplest yet important dynamical processes, which has been studied extensively on networks \cite{Noh2004,Hwang2012}. A key determinant of diffusion dynamics on networks is the spectral property of the Laplacian matrix, and in this sense diffusion dynamics is determined chiefly by the structure of networks. On this basis, there has been development of the mathematical formalism for studying diffusion processes on multiplex networks using the supra-Laplacian matrix~\cite{Gomez2013,Sole-Ribalta2013}. Suppose a simple $\ell$-plex network of $N$ nodes, in which particles diffuse with diffusion constant $D_{a}$ along each layers and $D_{ab}$ across different layers (Fig.~9a). 
Equations governing the time evolution of particle densities $x_i^{(a)}$ at each node $i$ on layer $a$ are given by 
\begin{align}
\frac{dx_{i}^{(a)}}{dt} = D_{a}\sum_{j=1}^{N}w_{ij}^{(a)}(x_{j}^{(a)}-x_{i}^{(a)}) + \sum_{\substack{b=1\\ b\neq a}}^{\ell}D_{ab}(x_{i}^{(b)}-x_{i}^{(a)}),
\end{align}
with $w_{ij}^{(a)}$ being the link-weight matrix of layer $a$, which can be recasted in matrix form as \cite{Gomez2013}
\begin{align}
\frac{d\mathbf{x}}{dt} = {\mathcal{L}}{\mathbf{x}}, 
\end{align} 
with $\mathbf{x}=(x_1^{(1)},\dots,x_N^{(1)},\dots,x_1^{(\ell)},\dots,x_N^{(\ell)})^{\rm T}$ and ${\cal L}$ being the supra-Laplacian matrix of the multiplex. 
For duplexes, the supra-Laplacian matrix can be written explicitly as
\begin{align}
\mathcal{L} = 
\left( \begin{array}{cc}
D_{A}L_{A} + D_{AB}I & -D_{AB}I \\
-D_{AB}I & D_{B}L_{B} + D_{AB}I
\end{array} \right),
\end{align}
where $L_{A}$ and $L_B$ are the Laplacian matrix of each layer, defined by $L_{a} = S_{a} - W_{a}$, where $W_{a}$ is the weights matrix of layer $a$ and $S_{a}$ is a diagonal matrix with the elements $(S_{a})_{ii} = \sum_{j}w_{ij}^{(a)}$~\cite{NewmanBook}. Of particular importance is the smallest nonzero eigenvalue $\lambda_2$ of the (supra) Laplacian matrix $\mathcal{L}$ which sets the diffusion timescale $\tau$ as $\tau=1/\lambda_{2}$, which characterizes how fast the system could relax to the stationary state (corresponding to $\lambda_1=0$). 

\begin{figure}
\centering
\includegraphics[width= .9\linewidth]{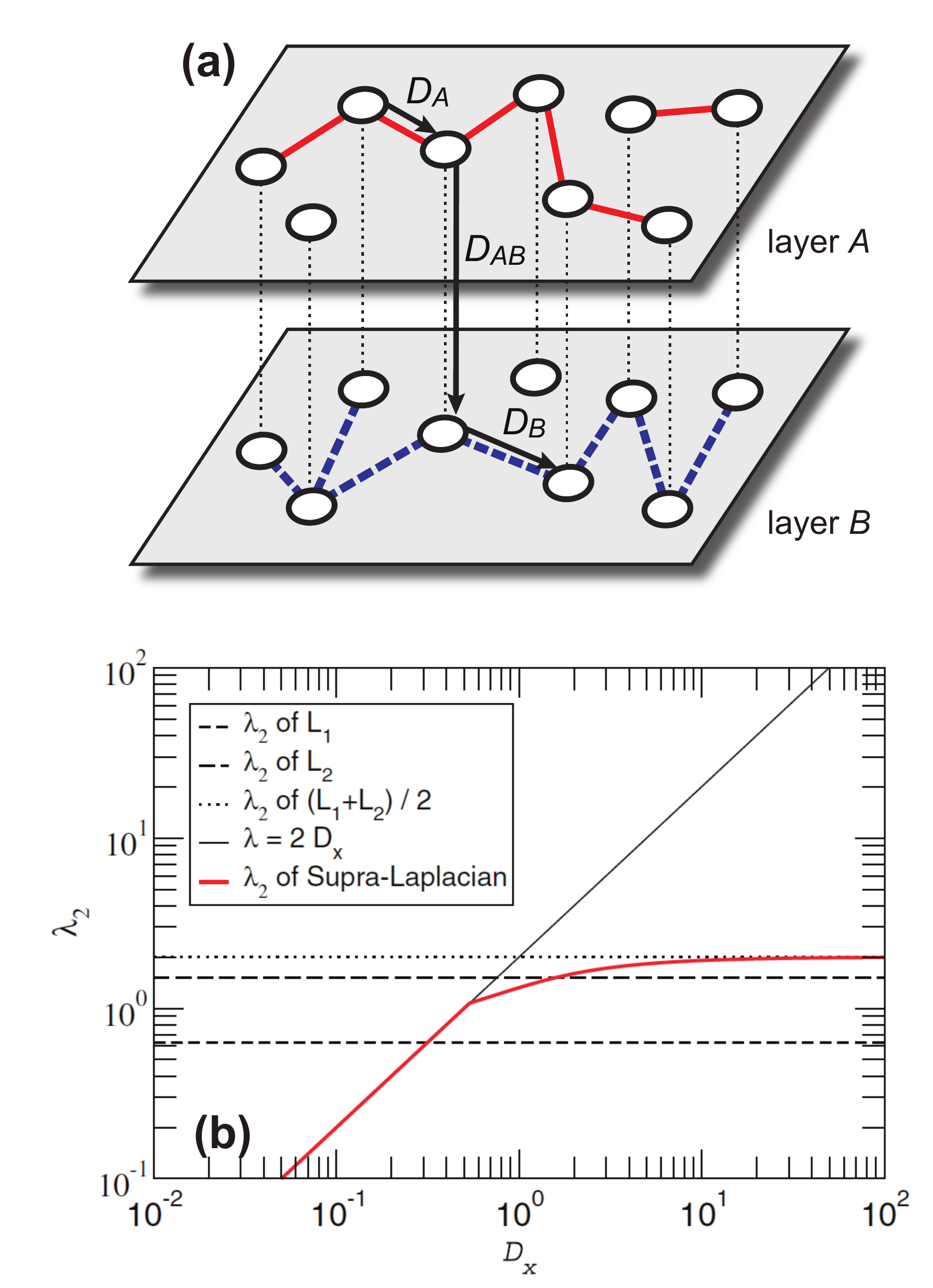}
\caption{(a) Schematic illustration for the diffusion process on a duplex network, with the intralayer diffusion constant, $D_A$ and $D_B$, and the interlayer diffusion constant, $D_{AB}$.
(b) Comparison between the second smallest eigenvalues $\lambda_2$ of the different Laplacians, as a function of the interlayer diffusion constant $D_{AB}$ (denoted as $D_x$ here) with $D_A=D_B=1$. Note that the diffusion timescale $\tau$ is given by $\tau=1/\lambda_2$. Panel (b) is reproduced from reference~\cite{Gomez2013} with permission from the authors and the publisher. Copyright (2013) by the American Physical Society.
}
\end{figure}

By investigating the eigenvalue spectrum of $\mathcal{L}$, equation~(28), a number of general conclusions have been drawn \cite{Gomez2013}, some of which we highlight below. 
First, the diffusion timescale $\tau$ or equivalently $\lambda_2$ undergoes a qualitative change at certain threshold interlayer diffusion strength $D_{AB}$ (Fig.~9b): if $D_{AB}$ is below the threshold, the diffusion timescale is completely governed by the interlayer diffusion, as $\tau=1/(2D_{AB})$. On the other hand, for $D_{AB}$ above the threshold, the timescale becomes dependent on the details of multiplex coupling, but it is always that the diffusion on the multiplex is faster than the diffusion in the slowest layer of the two. Only for sufficiently large $D_{AB}\gg1$, the diffusion on multiplex can become faster than the diffusion on any individual layer in isolation, referred to as the superdiffusive property in \cite{Gomez2013}\footnote{It is another misleading terminology, in that the term superdiffusion originally refers to the superlinear scaling of mean-square displacement in time in diffusive processes. Here the superdiffusive property  denotes merely an enhancement of effective diffusion constant, rather than a change in dynamic exponent. It is an intriguing open question whether scaling properties of diffusion dynamics and random walks such as spectral dimension \cite{Hwang2012} could be modified by multiplex coupling.}.
Further investigation of the supra-Laplacian matrices of interconnected networks  has identified the structural transition underlying the non-analyticity in diffusion timescale \cite{Radicchi2013}, which was later shown to be related with the reducibility transition of the supra-Laplacian matrices \cite{Garrahan2014}. 

Question of whether or not would it be possible to reduce a multiplex network into an equivalent single-layer network by layer aggregation was addressed also in terms of the supra-Laplacian matrix spectrum \cite{Domenico2014a}, in which it was claimed that for many cases the multiple layers of a multiplex could be reducible without significant loss of information. This result, however, may originate from the linear nature inherent to diffusion dynamics that the supra-Laplacian describes, and its applicability to general nonlinear dynamical problems is yet to be scrutinized. 
Meanwhile, as the Laplacian spectrum is closely related also with the stability of synchronized state \cite{Pecora1998,Nishikawa2003}, it was used in finding optimal coupling parameter for synchronizability in multiplex, interconnected systems~\cite{Sole-Ribalta2013}. 
Several other diffusion-related problems have been studied in the literature, such as the reaction-diffusion processes \cite{Xuan2013,Asllani2014}. 
Readers interested in further details are referred to a recent review on this subject  \cite{Salehi2014}. 

Non-diffusive transport such as the ballistic transport along the shortest paths can be relevant in such problems as the network routing and congestion control based on traffic load (or betweenness) \cite{Goh2001,Ahuja1993}. 
A pioneering work pointing out the relevance of multiple layers in such problems \cite{Kurant2006a} suggested the framework of what was called the layered network of the logical layer defined by the routing table and the physical layer over which the real transport occurs, and applied it to real traffic time table data of public transportation system to simulate real traffic patterns~\cite{Kurant2006b}. 
More recently, the multiplex framework has been applied in a more conceptually straightforward manner to the study of interconnected transportation systems~\cite{Morris2012}. Metropolitan transportation network is modeled as a coupled (interconnected) network of spatially-embedded layers. Depending on the origin-destination distribution, the optimal coupling strength may exist, for which the trade-off between time efficiency and congestion level can yield maximum utility of the system \cite{Morris2012}. The navigability of multiplex transportation system upon random failures was studied \cite{Domenico2014d} under both diffusive and shortest-path based transport scenarios.

\subsection{Epidemic spreading}
\label{subsec:epidemic}
Epidemic spreading is another classic topic of dynamical processes on networks with its own extensive literature~\cite{Pastor-Satorras2014}. It models how the infectious entities, be it a disease or a fad, spread over the network. The way the infection is transmitted to other nodes over  the links defines a specific model.
As the infection proceeds, it is not uncommon that agents (say individuals) in the network can be exposed to infectious entities (say human pathogens)  through many different channels of interaction (say different social network layers like family or co-workership) which can have vastly different spatiotemporal characteristics. 
To handle such scenario, the multiplex framework would not only be more straightforward conceptually but also more transparent methodologically than the single-layer counterpart, which in itself should be merely effective and approximate. 

Works on classical susceptible-infected-removed (SIR) and susceptible-infected-susceptible (SIS) epidemic spreading models on interconnected networks \cite{Dickison2012,Saumell-Mendiola2012,Yagan2013,Wang2013} should be noted as the predecessors of {\it bona fide} multiplex models. These models on interconnected networks could be studied by direct generalization of the single-layer methods, similarly to percolation and diffusion problems. 

Contagion process over multiplex social network with multiple channels (layers) of contagion as well as interlayer contagion following SIS-type dynamics was formulated by means of the contact-based discrete time Markov chain \cite{Cozzo2013}. It has shown that the epidemic threshold in such systems is completely governed by the layer with the largest maximum eigenvalue of the contact probability matrix and that simply aggregating different layers into a single network cannot describe the process accurately \cite{Cozzo2013}.
Analogous conclusion was drawn from the studies on similar models \cite{Vida2013,Buono2013}. 

\begin{figure}
\centering
\includegraphics[width= .81\linewidth]{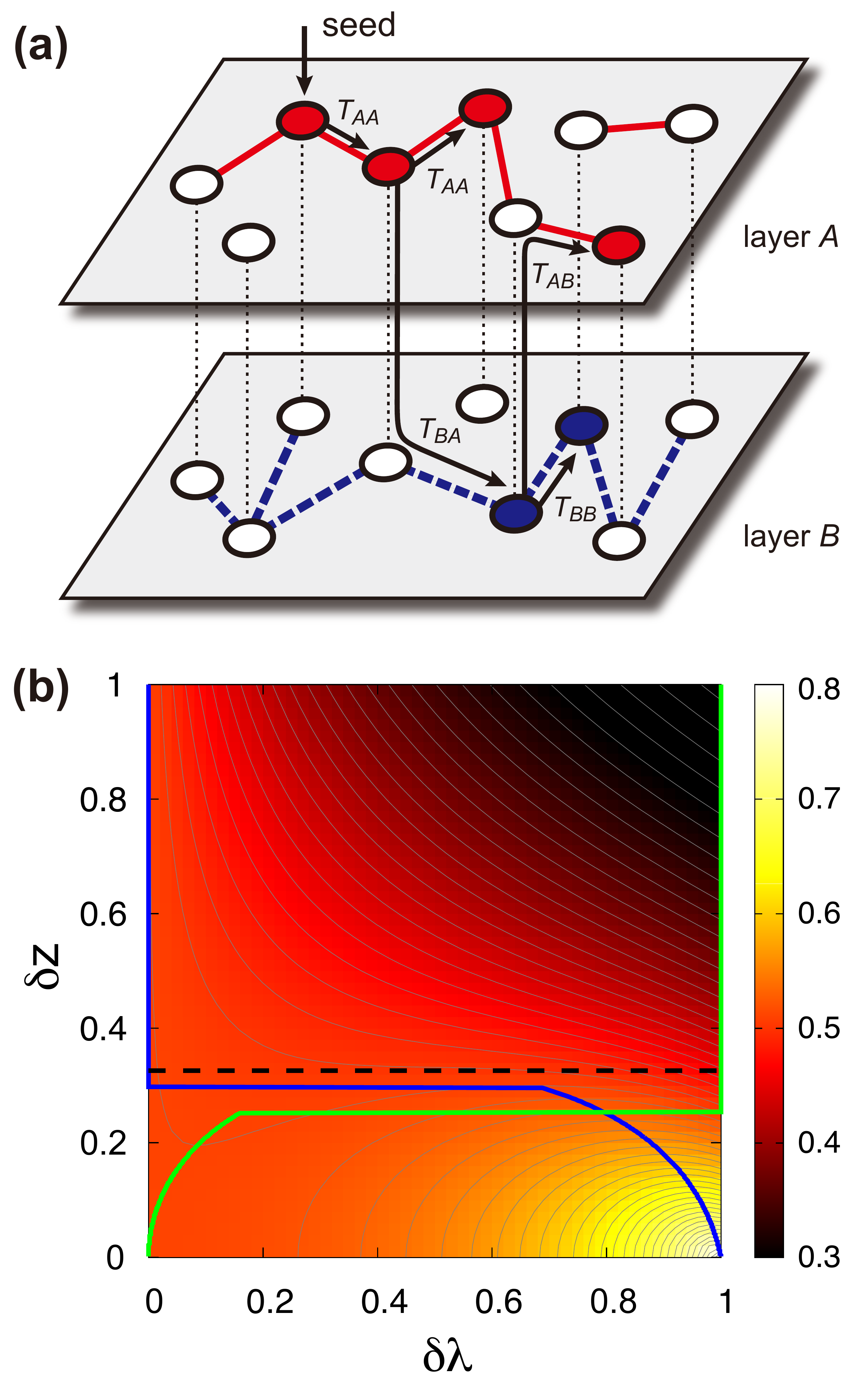}
\caption{(a) Schematic illustration of the SIR-type model with layer switching cost on a duplex network. Transmissibility $T$ for the infection along a link is determined by the types of both incoming (denoted by second subscript) and outgoing (first subscript) infection channels (layers). 
(b) Color-coded plot of the epidemic threshold $\lambda_c$ as a function of
the layer density disparity parameter $\delta z$, defined by $z_{A, B}=(1\pm\delta z)z_0/2$, 
and the infection rate disparity parameter $\delta\lambda$, defined by $\lambda_{\mathrm{intra, inter}}=(1\pm\delta\lambda){\lambda}$, on duplex ER networks with $z_0 = 2.5$. The horizontal dashed
line (black) is the boundary above which the epidemic threshold decreases monotonically with $\delta\lambda$. Green (blue) line indicates the loci with the lowest (highest) value of $\lambda_c$ for given $\delta z$, displaying non-analyticity. Adapted from reference~\cite{Min2013}.
}
\end{figure}
Often the coupled multiple layers contributing to spreading process are subject to some level of spatiotemporal separation. Consider the online and offline communication channels for information spreading, for instance. The concept of layer-crossing overhead (or layer-switching cost) was introduced as a new dynamic ingredient to account for this aspect in information spreading process over multiplex social networks  \cite{Min2013}. It implies the difference in the rates of across-the-layer and along-the-layer infections, leading to path-dependent transmissibility even for the same link (Fig.~10a). Studying an SIR-type model with such path-dependent transmissibility \cite{Min2013}, it was found that the layer-crossing overhead confers nontrivial effects on spreading dynamics, {\it viz.}, the epidemic threshold becomes dependent on the layer-crossing overhead in non-monotonic ways~(Fig.~10b). Novel analytical method tailored to deal with the path-dependent transmissibility had also to be developed \cite{Min2013}.  In this sense, it can be said that even the simple contagion model (like classical SIR model) can turn out not-so-simple on multiplex networks. 

Another situation for which multiplex framework is useful is the case of two or more infectious entities spreading through its own network layer over a multiplex population. Particularly interesting is when the multiple infectious entities interact, either compete or cooperate, for spreading.   The cases of competitive multiplex spreading were studied more actively, in the context of interplay between awareness and disease spreading \cite{Granell2013,Granell2014,WWang2014,Massaro2014}, interaction between two diseases coexisting in a host population such as AIDS and tuberculosis \cite{Sanz2014}, and spreading of competing memes \cite{Sahneh2014}, providing the conditions for coexistence or exclusive outbreak of different diseases in a population.
The coinfection model~\cite{Chen2013} is worthwhile to note even though it is studied on single-layer networks, as it provides a conceptual framework on how the cooperatively coupled epidemics (syndemics) could be modeled in multiplexes. It was found for the coinfection model the epidemic transitions can occur in a discontinuous manner together with hysteresis~\cite{Chen2013}, consonant with the findings from mutual percolation and viability problems in Section~3.

\section{Other types of layer coupling}

\subsection{Competitive coupling, antagonistic layers}
Another important type of layer couplings in multiplex systems is the competitive coupling of antagonistically interacting layers \cite{Zhao2013a,Aguirre2013}. It can also be regarded a subtype of cooperative coupling in that the layers are interlocking one another functionally, but we shall treat it separately to emphasize its unique features.

The concept of mutual percolation has been twisted into the percolation of antagonistic layers \cite{Zhao2013a,Zhao2013b} (which might well have been called the exclusive percolation\footnote{Not to be confused with the explosive percolation \cite{Achlioptas2009}.}). In this model, for two-layer multiplexes, the percolating cluster of each layer is defined in the following way: 
a node $i$ belongs to the percolating cluster of layer $A$ if it has at least one neighbor
in layer $A$ belonging to the percolating cluster of layer $A$, but at the same time, has no neighbors in layer $B$ belonging to the percolating cluster of layer $B$. ({\it vice versa} for layer $B$.)
The equations for the size of percolating clusters in each layer $S_a$ can be set up by following similar logic as equations~(8) and (9), as follows. First, the probability $w_{a}$ that the node reached by a randomly chosen $a$-layer link does not belong to the percolating cluster of layer $a$ satisfy the coupled self-consistency equations,
\begin{align}
1-w_{a} &=\sum_{\mathbf{k}}\frac{k_{a}P(\mathbf{k})}{z_{a}}(1-w_{a}^{k_{a}-1})w_{b}^{k_{b}}\nonumber\\
&= \left[1-G_{1}^{(a)}(w_a)\right]G_{0}^{(b)}(w_b),
\end{align}
for distinct $a, b\in\{1,2\}$, and the last equality holds when $P(\mathbf{k})$ factorizes (that is, the degrees of a node in different layers are uncorrelated). $G_0^{(a)}(x)$ and $G_1^{(a)}(x)$ are the generating functions defined as in equations~(21) and (22) but for each layer $a$.
Sizes of percolating clusters $S_a$ are given by \cite{Zhao2013a}
\begin{align}
S_{a} &=\sum_{\mathbf{k}}P(\mathbf{k})(1-w_{a}^{k_{a}})w_{b}^{k_{b}}\nonumber\\
&= \left[1-G_{0}^{(a)}(w_{a})\right]G_{0}^{(b)}(w_{b}),  
\end{align}
with the same convention as equation~(29).
Solving equations~(29) and (30) for a number of different layer topology, a rich phase diagram was obtained. Notably the bistable phase accompanying discontinuity and hysteresis in stable solutions was found, which occurs when the densities of both layers become sufficiently large~\cite{Zhao2013a} and sustains even with only a small fraction of antagonistic nodes \cite{Zhao2013b}.
These results might be compared with those found in epidemic spreading with competitive layers \cite{Granell2013,Granell2014,Sanz2014,Sahneh2014} for their similarity and differences (see Sect.~4.3). It is worth to remark that it has not yet been explicitly demonstrated how the different percolating clusters corresponding to solutions of equations~(29) and~(30) in the bistable phase could materialize from specific percolation processes, which is related to the question of the topological and algorithmic definition of the giant exclusively-connected component and its uniqueness. 
Finally, a \\conceptually-related study on two interconnected networks competing for spectral dominance~\cite{Aguirre2013} is worth a mention, in which successful interconnection strategy was investigated for a number of cases of different layer topology. 
Overall, compared to the previous two classes, much little attention has been paid to the competitively coupled layer problems thus far, remaining a fertile ground for future works. 

\subsection{Directional coupling, hierarchical layers}
Layers in a multiplex system can also have directional or hierarchical, rather than bidirectional or mutual, influence from one layer to another. Although this type of coupling has yet received little attention in the multiplex network literature, there is another active research field, the temporal network, in which the concept of directionally coupled layers is imposed naturally by the causality constraint due to time ordering. It is therefore expected that theoretical framework and understanding therefrom can be applicable to this type of multiplex problems \cite{Mucha2010}. The temporal network is itself an active branch of contemporary network theory, with growing body of literature. Readers interested in more details are referred to the recent comprehensive review and compendium on the subject \cite{Holme2012,HolmeBook}.

\section{Conclusion and outlook}
\label{conclusion}
In this Colloquium paper, we have presented an overview of recent development in multiplex networks from the viewpoint of statistical physics. In doing so, we have put consistent emphasis on the importance of bringing the proper context of multiplexity into the problem and its solution. Examples of multiplex systems and multiplex structural measures are introduced broadly yet concisely. Various statistical physics problems on the structure and dynamics of multiplex networks are categorized by the functional context of layer coupling. Two major categories, the problems with cooperative layers and those with complementary layers, are discussed in detail and highlighted for their distinction. 

No review papers can be completely exhaustive, and this paper is no exception. 
In a due manner, we have chosen deliberately to do a highly focused review centered on the strict definition of multiplex networks and the problems thereon. Closely-related variety of multilayered networks are drawn into in-depth discussion only when it is largely indispensable. Most of concepts and methodology can readily be applicable  to related multilayered networks with minimal customization. Readers are referred to~\cite{NoNBook,Kivela2014,Boccaletti2014} to fill the possible gap, however.
We have also limited the discussion for the most part on multiplex random networks for the sake of simplicity. 
Multiplexes of more heterogeneous layers such as scale-free networks could furnish a far richer repertoire of new behaviors due to the interplay between the structural singularity and multiplexity \cite{Buldyrev2010,Baxter2012,Jang2014}, a fertile ground for further study.
Last but not least, several other dynamical processes have also been studied in multiplex framework, detailed discussion of which was regretfully omitted in this paper. Notable examples include the Boolean dynamics on multilevel networks~\cite{Cozzo2012}, synchronization in interconnected networks~\cite{Um2011,Louzada2013,Aguirre2014,Nicosia2014c}, evolutionary game dynamics~\cite{Gomez-Gardenes2012a,Gomez-Gardenes2012b}, and opinion dynamics on multiplex coevolutionary networks \cite{Diakonova2014}. 

Despite some skepticism over the multiplex framework in favor of simpler and more-established single-layer one, the compelling viewpoint of this paper is that the multiplex framework is not only useful but crucial in elevating the level of our understanding of complex systems. Novel phenomena unforeseen in traditional single-layer framework can arise as a consequence of the coupling of network layers, especially when the layers are coupled in cooperative or competitive manner, in which cases the multiplex is not reducible into an equivalent single-layer system. It only remains to be seen what other new physics can emerge in this framework. In this prospect, useful lessons might be learned also from the vast literature on emergent phenomena in coupled statistical mechanical models~\cite{Ashkin1943,Selke1988,Galam1995}, as well as in layered physical systems such as layered cuprates in high-$T_c$ superconductors \cite{Anderson1997},  layered materials in multiferroics \cite{Cheong2007}, and layered graphenes~\cite{Geim2009}.

We would like to close this Colloquium paper with an idiosyncratic list of a few research questions warranting immediate attention. (i) Proper context-dependent definition and analysis of network measures such as the shortest path and betweenness centrality need to be established; (ii) There still remain major network concepts anticipating to be applied and generalized to multiplexes. A prime example is the controllability \cite{Liu2011,Nepusz2012,Luo2014,Yuan2014}; (iii)~No understanding in statistical physics is complete without systematic cataloguing and classification of possible universality and associated critical phenomena. Compared to single-network counterpart \cite{Goltsev2003}, the current level of our understanding is far from satisfactory. Identification of the minimal couplings (in the renormalization group sense) relevant to the characteristic discontinuous transitions in multiplex systems would be a seminal stepping stone in this regard.

\section*{Acknowledgements}
We thank Alex Arenas for his courtesy for Figure~9b. We would also like to thank Charles D. Brummitt, Won-kuk Cho, Jung Yeol Kim, and Sangchul Lee for productive collaboration on this exciting research topic. Special thanks goes to Jung Yeol Kim also for his help in collecting references. This work was supported by the Basic Science Research Program through an NRF grant funded by MSIP (Grant No. 2011-0014191).

\end{document}